\documentclass[hidelinks]{article}

\usepackage{arxiv}

\usepackage[utf8]{inputenc} 
\usepackage[T1]{fontenc}    
\usepackage{hyperref}       
\usepackage{url}            
\usepackage{booktabs}       
\usepackage{amsfonts}       
\usepackage{nicefrac}       
\usepackage{microtype}      
\usepackage{lipsum}		
\usepackage{graphicx}
\usepackage{natbib}
\usepackage{doi}
\usepackage{graphicx}
\usepackage{amsmath, amssymb}
\usepackage{color}
\usepackage[normalem]{ ulem }
\usepackage{soul}

\usepackage{bbold}
\usepackage{colortbl}
\usepackage{xcolor}

\title{Bayesian analysis of restricted mean survival time adjusted for covariates using pseudo-observations}

\author{ \href{https://orcid.org/0009-0007-7389-7612}{\includegraphics[scale=0.06]{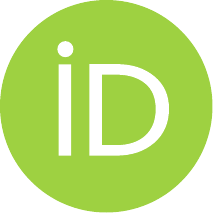}\hspace{1mm}Léa~Orsini} \\
	Oncostat U1018, Inserm\\
	University Paris-Saclay\\
	Villejuif, France \\
	\texttt{lea.orsini@gustaveroussy.fr} \\
	\And
	    \href{https://orcid.org/0000-0002-3747-6905}{\includegraphics[scale=0.06]{orcid.pdf}\hspace{1mm}Emmanuel~Lesaffre} \\
	    I-Biostat\\
        KU-Leuven\\
	    Leuven, Belgium \\
	    \texttt{emmanuel.lesaffre@kuleuven.be} \\
	    \And
	    \href{https://orcid.org/0000-0003-3276-1392}{\includegraphics[scale=0.06]{orcid.pdf}\hspace{1mm}Guosheng~Yin} \\
	    Department of Statistics and Actuarial Science\\
        The University of Hong Kong\\
        Hong Kong, China\\
	    \texttt{gyin@hku.hk} \\
	    \And
        \href{https://orcid.org/0000-0002-0196-4157}{\includegraphics[scale=0.06]{orcid.pdf}\hspace{1mm}Caroline~Brard} \\
	Ipsen Innovation, Clinical Development Organisation\\
	Les Ulis, France \\
	\texttt{caroline.brard@ipsen.com} \\
	    \And
	    \href{https://orcid.org/0000-0001-8960-5345}{\includegraphics[scale=0.06]{orcid.pdf}\hspace{1mm}David~Dejardin} \\
	    Product Development, Data Sciences\\
        F. Hoffmann-La Roche AG\\
	    Basel, Switzerland. \\
	    \texttt{david.dejardin@roche.com} \\
     \And
     	\href{https://orcid.org/0000-0003-3292-0939}{\includegraphics[scale=0.06]{orcid.pdf}\hspace{1mm}Gwénaël~Le Teuff} \\
        Oncostat U1018, Inserm\\
	    University Paris-Saclay\\
	    Villejuif, France \\
	    \texttt{gwenael.leteuff@gustaveroussy.fr} \\
}

\date{}

\renewcommand{\headeright}{}
\renewcommand{\undertitle}{}
\renewcommand{\shorttitle}{Bayesian analysis of restricted mean survival time adjusted for covariates using pseudo-observations}

\hypersetup{
pdftitle={Bayesian analysis of restricted mean survival time adjusted for covariates using pseudo-observations},
pdfauthor={Léa~Orsini, Emmanuel~Lesaffre, Guosheng~Yin, Caroline~Brard, David~Dejardin, Gwénaël~Le Teuff},
pdfkeywords={Bayesian survival analysis, Generalized method of moments, Non-proportional hazards, Pseudo-observations, Restricted mean survival time},
}

\begin{document}
\maketitle

\begin{abstract}
The difference in restricted mean survival time (RMST) is a clinically meaningful measure to quantify treatment effect in randomized controlled trials, especially when the proportional hazards assumption does not hold. Several frequentist methods exist to estimate RMST adjusted for covariates based on modeling and integrating the survival function. A more natural approach may be a regression model on RMST using pseudo-observations, which allows for a direct estimation without modeling the survival function. Only a few Bayesian methods exist, and each requires a model of the survival function. We developed a new Bayesian method that combines the use of pseudo-observations with the generalized method of moments. This offers RMST estimation adjusted for covariates without the need to model the survival function, making it more attractive than existing Bayesian methods. A simulation study was conducted with different time-dependent treatment effects (early, delayed, and crossing survival) and covariate effects, showing that our approach provides valid results, aligns with existing methods, and shows improved precision after covariate adjustment. For illustration, we applied our approach to a phase III trial in prostate cancer, providing estimates of the treatment effect on RMST, comparable to existing methods. In addition, our approach provided the effect of other covariates on RMST and determined the posterior probability of the difference in RMST exceeds any given time threshold for any covariate, allowing for nuanced and interpretable results.
\end{abstract}


\newpage
\section{Introduction}
\label{s:intro}

The difference in restricted mean survival time (dRMST) has been proposed as a clinically meaningful measure of the treatment effect of time-to-event outcomes in randomized controlled trials (RCTs), especially when the proportional hazards (PH) assumption does not hold since it does not rely on the PH assumption \citep*{Royston2011, CorroRamos2024} and provides a meaningful interpretation of the results \citep{Pak2017}. Non-PH occurs when the treatment effect is time-dependent (e.g., early or delayed effect). In recent years, a delayed treatment effect has often been observed when comparing immunotherapies with chemotherapies \citep{Anagnostou2017}. Non-PH may also occur with therapies that have different mechanisms of action. For example, treatment effects may be different between biomarker-positive and biomarker-negative groups (predictive biomarker), resulting in strongly different survival patterns. Even if the hazards are proportional within each group, non-PH can occur when considering the whole population, see for example \citet{Mok2009}. In these two settings, interpreting the hazard ratio as a treatment effect measure becomes tedious \citep{Hernan2010}. 

Extensive literature has been devoted to this measure \citep{Karrison1997, Royston2013, Hasegawa2020}. It is interpreted as the expected lifetime experienced out of $\tau$ units of time \citep{Irwin1949}. Suppose, for example, that dRMST at $\tau=5$ years between an experimental arm and a control arm in an RCT with an overall survival endpoint is estimated to be $1$ year. This means that, on average, the experimental treatment increases lifetime expectancy over the next $5$ years by $1$ year, compared to the control arm. 

One straightforward approach for estimating RMST at a certain time $\tau$ is to numerically integrate the Kaplan-Meier curve between 0 and $\tau$. However, this approach does not allow for covariate adjustment, which is a major limitation because omitting important covariates may result in less precision \citep{Karrison2018}. One way to adjust RMST estimation for covariates is to model the survival function with a parametric or semi-parametric model and integrate it, see \citet{Karrison1987} and \citet{Zucker1998}. A more natural approach is directly fitting a regression model on RMST through estimating equations. In this case, censoring must be handled using either the inverse probability of censoring weights \citep*{Tian2014} or pseudo-observations \citep*{Andersen2004}. 

RMST estimation can also be performed with a Bayesian approach, with a particular interest in rare diseases where small sample sizes are prevalent, and frequentist inference relying on asymptotic results may not be reliable \citep*{Lesaffre2020}. The Bayesian approach naturally allows the incorporation of prior information and may help to fit complex models. In addition, the Bayesian approach provides a more intuitive interpretation of the results by estimating the whole posterior distribution, giving an additional advantage compared to the frequentist RMST estimator. However, Bayesian research on RMST is limited. Recently, \citet{Zhang2023} proposed a Bayesian nonparametric approach to derive the RMST distribution from posterior samples for right and interval-censored data by assigning a Mixture of Dirichlet Processes (MDP) prior to the cumulative distribution function. However, this approach does not allow for covariate adjustment. \citet*{Chen2023} overcome this limitation by considering a data-adaptive stick-breaking prior based on a probit regression model. \citet{Hanada2024} estimate the RMST posterior distribution by fitting a parametric model on the survival function accounting for covariates and heterogeneity among clusters. All these methods require first modeling the survival function (or related functions) and then integrating them out in the two groups independently before subtracting the results, providing an indirect way of estimating the dRMST. 

In this paper, we extend the frequentist method of \citet{Andersen2004} based on the pseudo-observations to the Bayesian framework, offering a direct multivariable dRMST estimation without the need to model the survival function, thereby providing an attractive alternative to existing Bayesian methods.  

The rest of the paper is organized as follows. Section \ref{s:Methods} presents the Bayesian modeling of RMST using pseudo-observations with the Generalized Method of Moments (GMM). Section \ref{s:Simulation study} compares the performance of the GMM model to benchmark methods through a simulation study representing multiple scenarios of non-PH. We illustrate the methods through the analysis of a phase III RCT, the Getug-AFU 15 trial involving patients with prostate cancer, in Section \ref{s:application}. We conclude with some final remarks and future extensions of the proposed approach in Section \ref{s:discuss}.

\newpage
\section{Methods}
\label{s:Methods}
\subsection{RMST definition}
\label{ss:Definition}
Let $\rm \tilde{T}_i$ be the unobserved event time of interest $\rm (i=1,...,n)$ for the $\rm i$-th subject , $\rm Z_i$ a $\rm p$-dimensional baseline covariate vector, $\rm A_i$ the treatment allocation variable of an RCT ($\rm A_i$=1 and $0$ for experimental and control arm, respectively), and $\rm C_i$ a right censoring random variable, independent of $\rm \tilde{T}_i$, $\rm Z_i$ and $\rm A_i$. We observe $\rm T_i = \min(\tilde{T}_i,  C_i)$ and $\rm \Delta_i = I( \tilde{T}_i\leq C_i)$ the event indicator. For a pre-specified time point of interest $\tau$, the $\tau$-RMST is defined as
\begin{equation}\label{rmstdef}
\rm \rm RMST(\tau) = E(\min(\tilde{T}, \tau)) = \int_0^\tau S(t)dt.
\end{equation}

Under the specification of a generalized linear model, the subject-specific RMST at $\tau$ is defined by $\rm RMST_i(\tau)=E(\min(\tilde{T}_i, \tau))$ and can be conveniently modeled as
\begin{equation}\label{modelmu}
\rm \mu_i = E(\min(\tilde{T}_i, \tau)| A_i, Z_i) = g^{-1}(\beta_0 + \delta A_i + \beta_1 Z_{i1} + \cdots + \beta_p Z_{ip}),
\end{equation} where $\rm g(\cdot)$ is a monotone differentiable link function and $\rm \beta = (\beta_0, \delta, \beta_1, \ldots, \beta_p)^\top $ the vector of unknown parameters. For the identity link function, the regression coefficient $\delta$ can be interpreted as the dRMST between the two arms of an RCT. 

\subsection{Regression modeling of RMST using pseudo-observations}
\label{ss:RegressionPO}
In the frequentist framework, regression model (\ref{modelmu}) can be fitted using the pseudo-observations approach, see \citet*{Andersen2003}. Following \citet{Andersen2004}, the $i$-th pseudo-observation is computed as 
\begin{equation}\label{podef}
\rm y_{\tau,i} = n\int_0^\tau \widehat{S}(t)dt - (n-1)\int_0^\tau \widehat{S}^{-i}(t)dt,
\end{equation}
with $\rm n$ the sample size, $\rm \widehat{S}(t)$ the Kaplan-Meier (KM) estimator of the survival probability, and $\rm \widehat{S}^{-i}(t)$ the KM estimator excluding the $\rm i$-th subject. Given the asymptotic properties of pseudo-observations proved in \citet*{Overgaard2017}, we can replace the partially observed (due to censoring) $\rm \min(\tilde{T}_i, \tau)$ by $\rm y_{\tau,i}$ in regression model (\ref{modelmu}). Consequently, the pseudo-observations can be treated as outcome variables in a regression model to regress $\tau-$RMST on explanatory variables, using generalized estimating equations (GEE) \citep{Liang1986}. This marginal approach is based on quasi-likelihood functions where only the moments are defined \citep{McCullagh1991}.

The mean model is specified as:
\begin{equation}\label{modelmu2}
\rm{\mu_i = E(y_{\tau,i}| A_i, Z_i) = g^{-1}(\beta_0 + \delta A_i + \beta_1 Z_{i1} + \cdots + \beta_p Z_{ip})},
\end{equation}

The vector $\rm{\beta}$ is estimated by solving the score equations: 
    \begin{equation}\label{scorevec}
    \rm{U_n(\rm{\beta)}}= \frac{1}{n}\sum^n_{i = 1} \rm{\rm{u_i}(\rm{\beta)}}=\frac{1}{n}\sum^n_{i = 1} \frac{\partial \rm{\rm{\mu}_i}}{\partial\rm{\beta}}(\rm{y_{\tau,i}} - \rm{\rm{\mu}_i})=0,
    \end{equation}
    
\noindent where $\rm \frac{\partial \mu_i}{\partial\beta}$ is the first derivative of the mean model $\rm \mu_i$ with respect to $\rm \boldsymbol{\beta}$. In this case, the link function is the identity function.

\subsection{Bayesian generalized method of moments}
We propose to analyze pseudo-observations with a generalized linear model using the Generalized Method of Moments (GMM), defined by \citet{Hansen1982} in the frequentist framework and derived by \citet{Yin2009} in the Bayesian framework. The GMM is appropriate as it is a generic way of estimating parameters in statistical models, which is advantageous when the likelihood of the data is cannot be specified. The GMM does not require assuming a specific distribution for the data but is based on a number of moment conditions (first-order moment, and possibly higher-order moments). These moment conditions are combined through equations linking the model parameters and the data. If the number of equations exceeds the number of model parameters, the system of equations is over-identified, so the estimates are then found by minimizing an objective function.

To estimate RMST with the Bayesian GMM, with RMST model (\ref{modelmu}), pseudo-observations are computed as equation (\ref{podef}). Here, we are specifying only the first-order moment as in (\ref{modelmu2}). Therefore, the system of equations (\ref{scorevec}) holds. In this context, the objective function can be written as 
\begin{equation}\label{objectivefun}
\rm{Q_n(\rm{\beta}) = \rm{U_n}^\top\rm{(\rm{\beta})\Sigma_n}^{-1}\rm{(\rm{\beta})U_n(\rm{\beta})}},
\end{equation}
\noindent where $\rm{\rm{\Sigma}_n}(\rm{\beta}) =  \frac{1}{n^2}\sum_{i = 1}^n \rm{u_i(\rm{\beta})}\rm{u_i}^\top(\rm{\beta}) - \frac{1}{n}\rm{U_n(\rm{\beta})}\rm{U_n}^\top(\rm{\beta})$. 

\noindent By applying the Central Limit Theorem, 
\begin{equation}\label{Normallimit}
\rm{U_n(\rm{\beta})} \overset{d}{\longrightarrow} N(0,\rm{\Sigma}(\rm{\beta})), \text{as $\rm n\to \infty$}
\end{equation}

\noindent where $\rm{\Sigma}(\rm{\beta}) = \underset{n \to \infty}{\lim}\rm{\rm{\Sigma}_n}(\rm{\beta}),$ then
\begin{equation}\label{Khisquarelimit}
\rm Q_n(\rm{\beta}) \overset{d}{\longrightarrow} \chi_{p+2}^2.
\end{equation}
A chi-squared test can be derived where $\rm Q_n(\rm{\beta})$ behaves like $\rm -2\log L(\rm \beta | y)$ with $\rm L(\rm \beta | y)$ the likelihood function \citep{Hansen1982}. Thus, the objective function can be seen as an approximation of the likelihood for selected moments of the data when the full likelihood of the data is unknown, \citep{Chernozhukov2003}. A Bayesian sampling procedure can be developed where the likelihood function is approximated by defining a pseudo-likelihood function as follows: 
\begin{equation}\label{pseudolik}
\rm \tilde{L}(\rm \beta | y) \propto \exp\{-\frac{1}{2}\rm{U_n}^\top(\rm{\beta})\rm{\rm{\Sigma}_n}^{-1}(\rm{\beta})\rm{U_n(\rm{\beta})}\}.
\end{equation}
\citet{Yin2009} demonstrated the validity of the posterior distribution sampled with this pseudo-likelihood. The posterior distribution for the regression coefficient, $\delta,$ represents the posterior distribution for the dRMST between the two treatment arms.

\section{Simulation study}
\label{s:Simulation study}
A simulation study of 2-arms RCTs was conducted to assess the performance of the new Bayesian approach based on pseudo-observations for estimating RMST. The purpose of this simulation study was to (a) assess the validity of the dRMST estimation using the Bayesian GMM based on pseudo-observations, (b) compare these estimates with other frequentist and Bayesian dRMST estimates (benchmark methods), and (c) evaluate the impact of covariates adjustment and model misspecification.

\subsection{Simulation settings}
\begin{figure}[!ht]\centering
\includegraphics[width=16.5cm]{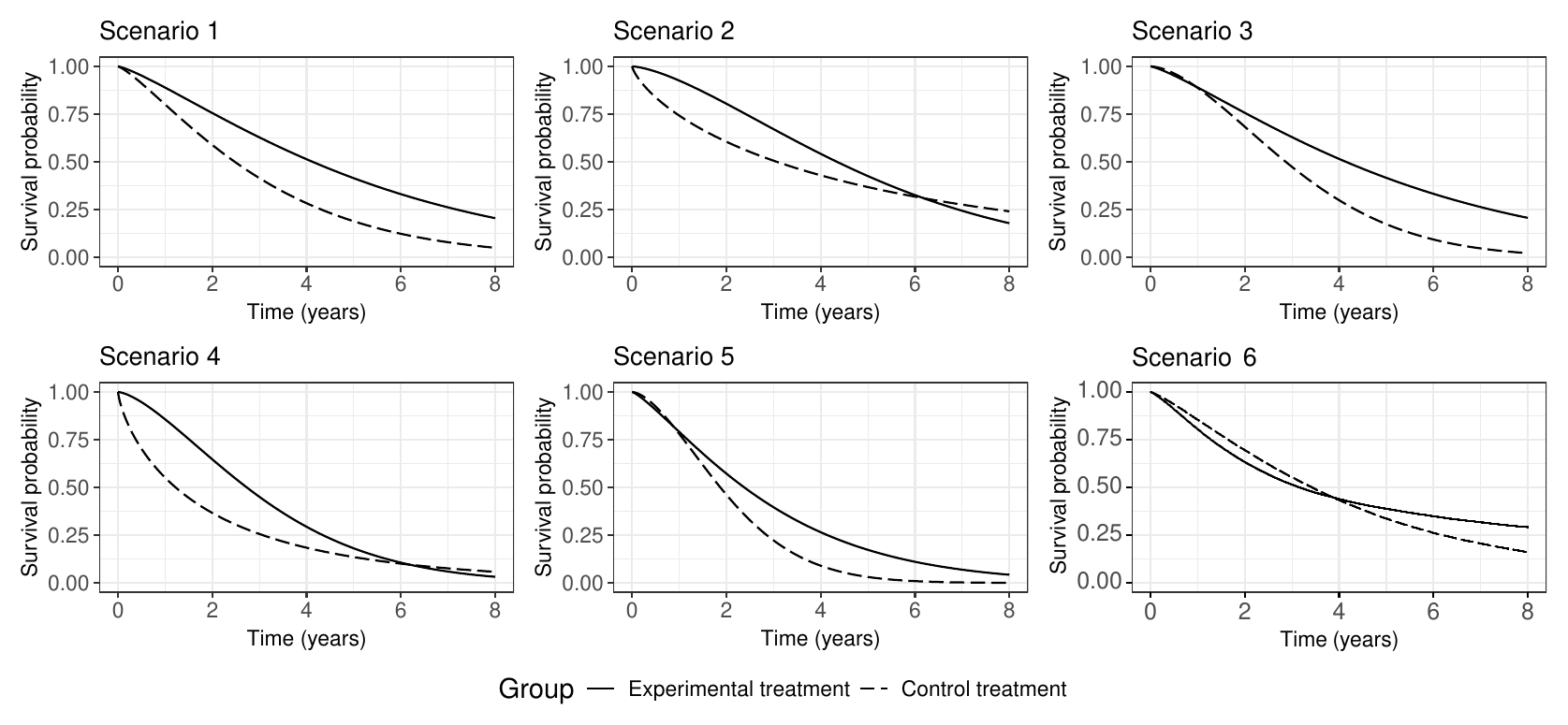}
\caption{True survival curves for the different scenarios.}
\label{orsini:fig1}
\end{figure}

 We simulated individual survival data of a 2-arm RCT (ratio 1:1), for different scenarios of the treatment effect (experimental vs control) over time with PH (scenario 1) and non-PH (scenarios 2-6). Figure \ref{orsini:fig1} shows the true survival curves for each scenario. In scenarios 2 and 4, we simulate an early treatment effect, in scenarios 3 and 5, we simulate a delayed treatment effect, and in scenario 6, we simulate crossing survival curves. In scenarios 4 and 5, we simulate survival data associated with prognostic variables. These two scenarios allow us to evaluate the impact of covariate adjustments and model misspecification when omitting important prognostic variables. For each scenario, the event times were simulated according to a Weibull distribution $\rm S(t) = \exp(-(\lambda t)^{1/\sigma})$, with scale $\lambda$ and shape $\sigma$ parameters chosen to mimic different patterns of treatment effect. The corresponding true value of the RMST is 
 \begin{equation}\label{true_dRMST}
 \rm  \int_0^\tau \exp(-(\lambda t)^{1/\sigma})dt = \frac{\sigma}{\lambda}\gamma(\sigma, (\lambda\tau)^{1/\sigma}),
 \end{equation} 
 
\noindent where $\rm \gamma(\rm a, x) = \int_0^{x}t^{\rm a-1}e^{-t}dt$ is the lower incomplete gamma function. The true value of the dRMST ($\delta$) can be calculated by subtracting the true RMST values of the two arms. The censoring times were drawn from a uniform distribution, with an additional administrative censoring at $8$ years, yielding an average of $30\%$ of censored individuals per scenario. Prognostic covariates $\rm Z_{ij}$ were drawn from a uniform distribution (scenario 4) or normal and Bernoulli distributions (scenario 5). Their effects on time-to-event satisfy the proportional hazards assumption. Other variables $\rm X_{ij}$, unrelated to the survival function, were also simulated from a normal, a Bernoulli, and a uniform distribution, respectively, to assess the impact of covariate adjustments in terms of model misspecification in scenarios 4 and 5. In scenario 6, crossing survival curves between the experimental and control arms were simulated by including an interaction between the treatment indicator and a binary variable $\rm E_i$ with a predictive value. Consequently, the treatment effect depends on whether the biomarker takes $\rm E_i = E^-$ or $\rm E_i = E^+$. We assume that the treatment was harmful if $\rm E_i = E^-$, and the treatment was protective if $\rm E_i = E^+$. The different simulated scenarios are summarized in supporting information Table 1. For simplicity, we will omit the $\rm i$ index in $\rm A_i$, $\rm Z_{ij}$, $\rm X_{ij}$, and $\rm E_i$ throughout the rest of this paper. Different sample sizes of simulated RCTs were considered, i.e. $50$, $100$, $200$, and $500$. According to the follow-up of the simulated RCTs (accrual period and administrative censoring), RMST was estimated at the restriction time $\tau$ set to $5$ years, often used as a clinically meaningful threshold in oncology trials \citep{Chen2023}. However, when the last simulated observed time in one arm was smaller than 5 years, then $\tau$ was redefined as the minimum of the maximum observed times of each arm. The number of times that this situation occurs is reported in supporting information Table 2.

The performance of the Bayesian GMM and the benchmark methods (See subsection \ref{s:Benchmarkmethods} for a description of these methods) was evaluated using the following metrics calculated from $1000$ replications: bias, average standard errors (ASE: the mean of the estimated standard errors of $\widehat{\delta}$ across replications), empirical standard errors (ESE: the standard deviation of $\widehat{\delta}$ across replications), root mean square error (RMSE = $\rm \sqrt{ESE^2 + Bias^2}$), and the coverage rate of the intervals (either $95\%$ confidence intervals or equal-tailed credible intervals for the frequentist and the Bayesian approaches, respectively). The convergence of the Bayesian procedures was checked with trace plots checking and Gelman-Rubin $\rm \widehat{R}$ diagnostic \citep{Vehtari2021}.

All computations were performed using the R Language for Statistical Computing  (\citet{RCoreTeam2021}, version 4.1.2). Pseudo-observations have been computed using the R package \texttt{pseudo} \citep*{PoharPerme2017}. We developed a specific script to implement the Bayesian GMM using the Stan software \citep{Carpenter2017}. The Hamiltonian Monte Carlo algorithm was performed using the No-U-turn (NUTS) sampler with $3$ chains of each $1000$ iterations after a warm-up of $1000$ iterations, yielding $3000$ iterations overall. Weakly informative priors were specified using a normal distribution $\rm N(\mu = 0, \sigma^2 = 10)$ for all parameters of the Bayesian GMM with pseudo-observations, as recommended by the \citet{StanDev2020}. Regarding computational time, for one run of the Bayesian GMM with $\rm n = 200$ patients and $3$ explanatory variables included in the model, the pseudo-observations computation took less than $1$ second, and the running time of one chain for the Bayesian GMM was $10$ seconds with an 11th Gen Intel(R) Core(TM) i5-1135G7 processor. Using the \texttt{rstan} R package \citep{StanDev2023}, chains can run in parallel, not adding additional running time when increasing the number of chains. So, for the simulation study, the Bayesian GMM was run with a parallelized code using a server HPE DL385 (2.0 GHz) with $150$ virtual cores due to the high numbers of scenarios and replicates per scenario. The median running time of one chain according to different sample sizes and numbers of explanatory variables included in the model is provided in supporting information Figure 1 for scenario 5. Regarding Bayesian GMM convergence, only scenarios 4 and 5, which involve covariate adjustment with prognostic and unrelated variables, showed a small percentage of non-convergent replications (less than $1.5\%$, supporting information Table 3), which were subsequently removed from the analysis.

\subsection{Benchmark methods}
\label{s:Benchmarkmethods}
Four benchmark methods, providing an estimator of dRMST, have been retained for comparison: three frequentist estimators and one Bayesian. The first one (naive approach) consists of integrating the Kaplan-Meier estimator in both arms separately before subtracting the estimated RMST values of the two arms. The second consists of applying a Cox model stratified on treatment \citep{Cox1972}, with \citet{Breslow1972} estimator to estimate the baseline hazard function in each arm and model the survival function conditional on covariates before averaging over the covariates, integrating it and subtracting the estimated RMST values of the two arms, see \citet{Zucker1998} and the third benchmark method is the pseudo-observations approach from \citet{Andersen2004} that proposed GEE to model RMST. The R package \texttt{survival} was used to estimate the RMST from the Kaplan-Meier estimator \citep{Therneau2024}. The R package \texttt{geepack} was used to implement the GEE approach for fitting marginal generalized linear model to pseudo-observations \citep{Højsgaard2022}. The Bayesian benchmark method is the one of \citet{Zhang2023} who proposed a nonparametric approach for estimating the RMST by independently assigning an MDP prior to the cumulative incidence function of each arm. An MDP prior was applied under an exponential base measure with a Gamma $\Gamma (0.01, 0.01)$ mixing distribution.

\subsection{Unadjusted estimation of the RMST}
In scenario 1, the simulation parameters were set to $\sigma = 0.8$ and $\rm \lambda= \exp(-1.2 + \log(HR)\rm{A_i})$ expressing the treatment effect as a hazard ratio $\rm HR$. Under the null hypothesis with proportional hazards (no treatment effect, $\rm HR = 1$ thus $\delta=0$), the Bayesian GMM produced similar estimations of the dRMST between the two arms compared to the four benchmark methods: no bias, ASE and ESE were close and decrease when sample size increased and coverage was close to the nominal $95\%$ rate when the sample size was large enough ($\rm n \geq 100$) (see supporting information Table 4). More generally, the performance of the different methods became similar when the sample size increased. Under an alternative hypothesis ($\sigma=0.8$ and the treatment effect $\rm HR=0.6$, resulting in $\delta=0.82$ when applying the formula (\ref{true_dRMST})), again the results were similar to the null hypothesis setting (see supporting information Table 5). 

In scenarios 2 and 3, we simulated a treatment effect with non-proportional hazards by specifying different scale and shape parameters for each treatment group. In scenario 2, we set $\sigma_0 = 1.33$, $\lambda_0 = 0.20$ for the control group and $\sigma_1 = 0.67$, and $\lambda_1 = 0.18$ for the treatment group, resulting in survival curves presenting an early treatment effect that decreases through the years. In scenario 3, we set $\sigma_0 = 0.60$, $\lambda_0 = 0.28$ for the control group and $\sigma_1 = 0.80$, $\lambda_1 = 0.18$ for the treatment group, resulting in survival curves presenting a delayed treatment effect that increases through the years. Table~\ref{orsini:table1} reports the results for these 2 scenarios with sample sizes of n=50, 100, and 200 (See supporting information Table 6 and supporting information Table 7 for an additional sample size of n=500). The findings were similar to what we observed for scenario 1, i.e., the Bayesian GMM gave a valid estimate of the dRMST for any sample size with performance closer to those of the four benchmark methods as the sample size increases. For large sample sizes ($\rm n = 500$), the RMSE closely approximated the ESE. Conversely, for $\rm n= 50$, in scenarios 1-3, we note that the Bayesian GMM was more conservative with a higher ASE compared to the four benchmark methods, resulting in a coverage rate closer to the $95\%$ and around $92\%-93\%$ for the other methods.

\begin{table}[!ht]\centering
\caption{Performance of the frequentist and Bayesian methods on the univariate estimation of the difference of restricted mean survival time (dRMST) between two treatment groups with non-proportional hazards treatment effect (scenario 2: early effect $\delta = 0.7302$, scenario 3: late effect $\delta = 0.5644$), with $30\%$ of censoring and different sample sizes.}
\label{orsini:table1}
\medskip
\begin{tabular}{cllrrrrr}
\toprule
Scenario & n & Methods & Bias & ASE$^1$ & ESE$^2$ & RMSE$^3$ & Coverage\\
\midrule
\addlinespace[0.3em]
2&50&\multicolumn{5}{l}{\textbf{Frequentist}}\\
&&\hspace{1em}KM estimator & -0.0056 & 0.498 & 0.538 & 0.538 & 92.2\\
&&\hspace{1em}Zucker (1998) & -0.0198 & 0.493 & 0.526 & 0.526 & 92.3\\
&&\hspace{1em}Andersen et al. (2004) & -0.0058 & 0.499 & 0.538 & 0.538 & 92.1\\
\addlinespace[0.3em]
&&\multicolumn{6}{l}{\textbf{Bayesian}}\\
&&\hspace{1em}Zhang and Yin (2023) & -0.0061 & 0.487 & 0.538 & 0.538 & 92.1\\
&&\hspace{1em}GMM & 0.0213 & 0.514 & 0.521 & 0.522 & 93.2\\
\cmidrule(rl){2-8}
\addlinespace[0.3em]
 &100&\multicolumn{5}{l}{\textbf{Frequentist}}\\
&&\hspace{1em}KM estimator & 0.0068 & 0.356 & 0.356 & 0.356 & 94.3\\
&&\hspace{1em}\citet{Zucker1998} & -0.0006 & 0.354 & 0.352 & 0.352 & 94.6\\
&&\hspace{1em}\citet{Andersen2004} & 0.0068 & 0.356 & 0.356 & 0.356 & 94.4\\
\addlinespace[0.3em]
&&\multicolumn{6}{l}{\textbf{Bayesian}}\\
&&\hspace{1em}\citet{Zhang2023} & 0.0067 & 0.352 & 0.356 & 0.356 & 94.4\\
&&\hspace{1em}GMM & 0.0192 & 0.359 & 0.351 & 0.351 & 94.8\\
\cmidrule(rl){2-8}
\addlinespace[0.3em]
&200&\multicolumn{5}{l}{\textbf{Frequentist}}\\
&&\hspace{1em}KM estimator & -0.0071 & 0.252 & 0.254 & 0.254 & 95.4\\
&&\hspace{1em}\citet{Zucker1998} & -0.0107 & 0.251 & 0.252 & 0.253 & 95.3\\
&&\hspace{1em}\citet{Andersen2004} & -0.0070 & 0.252 & 0.254 & 0.254 & 95.4\\
\addlinespace[0.3em]
&&\multicolumn{6}{l}{\textbf{Bayesian}}\\
&&\hspace{1em}\citet{Zhang2023} & -0.0072 & 0.251 & 0.254 & 0.254 & 95.3\\
&&\hspace{1em}GMM & -0.0020 & 0.252 & 0.252 & 0.252 & 95.4\\
\midrule
\addlinespace[0.3em]
3& 50&\multicolumn{5}{l}{\textbf{Frequentist}}\\
&&\hspace{1em}KM estimator & -0.0212 & 0.461 & 0.491 & 0.492 & 92.7\\
&&\hspace{1em}Zucker (1998) & -0.0278 & 0.459 & 0.485 & 0.486 & 93.1\\
&&\hspace{1em}Andersen et al. (2004) & -0.0206 & 0.464 & 0.492 & 0.492 & 93.1\\\addlinespace[0.3em]
&&\multicolumn{6}{l}{\textbf{Bayesian}}\\
&&\hspace{1em}Zhang and Yin (2023) & -0.0129 & 0.451 & 0.497 & 0.497 & 92.3\\
&&\hspace{1em}GMM & -0.0044 & 0.482 & 0.480 & 0.480 & 94.4\\
\cmidrule(rl){2-8}
\addlinespace[0.3em]
&100&\multicolumn{5}{l}{\textbf{Frequentist}}\\
&&\hspace{1em}KM estimator & 0.0058 & 0.330 & 0.340 & 0.340 & 93.5\\
&&\hspace{1em}\citet{Zucker1998} & -0.0016 & 0.329 & 0.336 & 0.336 & 93.9\\
&&\hspace{1em}\citet{Andersen2004} & 0.0055 & 0.331 & 0.339 & 0.339 & 93.7\\
\addlinespace[0.3em]
&&\multicolumn{6}{l}{\textbf{Bayesian}}\\
&&\hspace{1em}\citet{Zhang2023} & 0.0059 & 0.326 & 0.340 & 0.340 & 93.4\\
&&\hspace{1em}GMM & 0.0124 & 0.336 & 0.335 & 0.335 & 94.3\\
\cmidrule(rl){2-8}
\addlinespace[0.3em]
&200&\multicolumn{5}{l}{\textbf{Frequentist}}\\
&&\hspace{1em}KM estimator & -0.0065 & 0.234 & 0.237 & 0.237 & 93.9\\
&&\hspace{1em}\citet{Zucker1998} & -0.0101 & 0.233 & 0.235 & 0.236 & 94.0\\
&&\hspace{1em}\citet{Andersen2004} & -0.0063 & 0.235 & 0.237 & 0.237 & 93.9\\
\addlinespace[0.3em]
&&\multicolumn{6}{l}{\textbf{Bayesian}}\\
&&\hspace{1em}\citet{Zhang2023} & -0.0065 & 0.232 & 0.237 & 0.237 & 94.1\\
&&\hspace{1em}GMM & -0.0035 & 0.236 & 0.236 & 0.236 & 95.1\\
\bottomrule
\multicolumn{8}{l}{\rule{0pt}{1em}\textsuperscript{1}ASE: Average Standard Error, \textsuperscript{2}ESE: Empirical Standard Error, \textsuperscript{3}RMSE: Root Mean Square Error}\\
\end{tabular}
\vspace{1em}
\end{table}

\subsection{Adjusted estimation of the RMST}
In scenarios 4 and 5, we simulated situations where RMST was associated with one prognostic variable $\rm Z_1$ (scenario 4) and two prognostic variables $\rm Z_1$ and $\rm Z_2$ (scenario 5). In scenario 4, we simulated an early treatment effect with $\rm  Z_1 \sim U(0, 2)$. In scenario 5, we simulated a delayed treatment effect with $\rm  Z_1 \sim N(0, 1)$ and $\rm  Z_2 \sim Bin(0.5)$. Table \ref{orsini:tab2} reports the results with and without covariates-adjustment for an RCT of $\rm n = 200$. This corresponds to the first two and the first three rows of each method (\citet{Zucker1998, Andersen2004} and Bayesian GMM) for scenario 4 and scenario 5, respectively. All methods allowing for adjustment for covariates produced slightly more precise estimates when the model was well specified compared to the misspecified univariate analysis with a relative gain in precision from unadjusted to adjusted estimations of $4\%$ (scenario 4) and $8\%$ (scenario 5) for the Bayesian GMM. Omitting prognostic variables also produced estimates with less precision with $\rm n = 100$ and $\rm n = 500$, see supporting information Table 8. For a smaller sample size ($\rm n = 50$), we did not observe an increase in precision with the Bayesian GMM when adjusting for the prognostic variables, while the frequentist models of \citet{Zucker1998} and \citet{Andersen2004} still provided smaller ASE and satisfactory coverage. This behavior was still reasonably small when adjusting for one covariate: in scenario 4, an ASE of $0.517$ ($94.5\%$ coverage rate) for the unadjusted model compared to $0.532$ ($95.5\%$ coverage rate) when adjusting for $\rm  Z_1$. When we adjusted for two covariates, the variance increased more: in scenario 5, ASE of $0.517$ ($96.2\%$ coverage rate) for the unadjusted model compared to an ASE of $0.742$ ($99.1\%$ coverage rate) when adjusting for $\rm  Z_1$ and $\rm  Z_2$. On the other hand, the ESE of the Bayesian GMM stayed comparable to the ones of the other approaches, see supporting information Table 9. 

\begin{table}[!ht]\centering
\caption{\label{orsini:tab2} Performance of the frequentist and Bayesian methods on the estimation of the difference of restricted mean survival time (dRMST) between two treatment groups, adjusted for prognostic covariates $\rm  Z_j$ and non-prognostic covariates $\rm  X_j$, with non-proportional hazards treatment effect (scenario 4: early effect $\delta = 0.9532$, scenario 5: late effect $\delta = 0.4911$), with $30\%$ of censoring and a sample size of $200$. The gray shaded lines represent the correctly specified model.}
\medskip
\begin{tabular}{cllrrrrr}
\toprule
Scenario & Methods & Adjustment variables & Bias & ASE$^1$ & ESE$^2$ & RMSE$^3$ & Coverage\\
\midrule
\addlinespace[0.3em]
4 &\multicolumn{5}{l}{\textbf{Frequentist}}\\
&\hspace{1em}KM estimator & - & -0.0056 & 0.257 & 0.266 & 0.266 & 93.8\\
&\hspace{1em}\citet{Zucker1998} & - & -0.0104 & 0.258 & 0.264 & 0.264 & 93.8\\
\rowcolor{gray!25}&\hspace{1em}\citet{Zucker1998} & $\rm  Z_1$ & -0.0133 & 0.239 & 0.243 & 0.243 & 93.9\\
&\hspace{1em}\citet{Zucker1998} & $\rm  Z_1$, $\rm  X_1$ & -0.0128 & 0.239 & 0.245 & 0.245 & 93.9\\
&\hspace{1em}\citet{Zucker1998} & $\rm  Z_1$, $\rm  X_1$, $\rm  X_2$, $\rm  X_3$ & -0.0117 & 0.239 & 0.248 & 0.248 & 93.8\\
&\hspace{1em}\citet{Andersen2004} & - & -0.0056 & 0.258 & 0.266 & 0.266 & 93.8\\
\rowcolor{gray!25}&\hspace{1em}\citet{Andersen2004} & $\rm  Z_1$  & -0.0088 & 0.246 & 0.251 & 0.251 & 93.9\\
&\hspace{1em}\citet{Andersen2004} & $\rm  Z_1$, $\rm  X_1$ & -0.0074 & 0.246 & 0.251 & 0.251 & 93.9\\
&\hspace{1em}\citet{Andersen2004} & $\rm  Z_1$, $\rm  X_1$, $\rm  X_2$, $\rm  X_3$ & -0.0068 & 0.246 & 0.254 & 0.254 & 93.7\\
\addlinespace[0.3em]
&\multicolumn{7}{l}{\textbf{Bayesian}}\\
&\hspace{1em}\citet{Zhang2023} & - & -0.0058 & 0.256 & 0.266 & 0.266 & 93.8\\
&\hspace{1em}GMM & - & -0.0070 & 0.259 & 0.263 & 0.263 & 94.5\\
\rowcolor{gray!25}&\hspace{1em}GMM & $\rm  Z_1$ & -0.0033 & 0.250 & 0.249 & 0.249 & 94.6\\
&\hspace{1em}GMM & $\rm  Z_1$, $\rm  X_1$ & -0.0014 & 0.253 & 0.249 & 0.249 & 94.3\\
&\hspace{1em}GMM & $\rm  Z_1$, $\rm  X_1$, $\rm  X_2$, $\rm  X_3$ & -0.0008 & 0.261 & 0.252 & 0.252 & 95.3\\
\midrule
\addlinespace[0.3em]
5 & \multicolumn{5}{l}{\textbf{Frequentist}}\\
&\hspace{1em}KM estimator & - & -0.0062 & 0.251 & 0.262 & 0.263 & 93.5\\
&\hspace{1em}\citet{Zucker1998} & - & -0.0098 & 0.251 & 0.261 & 0.261 & 93.5\\
&\hspace{1em}\citet{Zucker1998} & $\rm  Z_1$ & -0.0024 & 0.215 & 0.223 & 0.223 & 94.6\\
\rowcolor{gray!25}&\hspace{1em}\citet{Zucker1998} & $\rm  Z_1$, $\rm  Z_2$ & -0.0031 & 0.211 & 0.219 & 0.219 & 94.7\\
&\hspace{1em}\citet{Zucker1998} & $\rm  Z_1$, $\rm  Z_2$, $\rm  X_1$, $\rm  X_2$& -0.0031 & 0.211 & 0.220 & 0.220 & 94.0\\
&\hspace{1em}\citet{Andersen2004} & - & -0.0061 & 0.252 & 0.263 & 0.263 & 93.5\\
&\hspace{1em}\citet{Andersen2004} & $\rm  Z_1$ & -0.0025 & 0.227 & 0.235 & 0.235 & 93.8\\
\rowcolor{gray!25}&\hspace{1em}\citet{Andersen2004} & $\rm  Z_1$, $\rm  Z_2$ & -0.0045 & 0.225 & 0.233 & 0.233 & 93.9\\
&\hspace{1em}\citet{Andersen2004} & $\rm  Z_1$, $\rm  Z_2$, $\rm  X_1$, $\rm  X_2$& -0.0039 & 0.225 & 0.234 & 0.234 & 93.8\\
\addlinespace[0.3em]
&\multicolumn{6}{l}{\textbf{Bayesian}}\\
&\hspace{1em}\citet{Zhang2023} & - & -0.0061 & 0.250 & 0.262 & 0.263 & 93.6\\
&\hspace{1em}GMM & - & -0.0029 & 0.253 & 0.261 & 0.261 & 94.2\\
&\hspace{1em}GMM & $\rm  Z_1$ & 0.0046 & 0.233 & 0.234 & 0.234 & 94.8\\
\rowcolor{gray!25}&\hspace{1em}GMM & $\rm  Z_1$, $\rm  Z_2$ & 0.0021 & 0.232 & 0.232 & 0.232 & 94.7\\
&\hspace{1em}GMM & $\rm  Z_1$, $\rm  Z_2$, $\rm  X_1$, $\rm  X_2$& 0.0032 & 0.239 & 0.234 & 0.234 & 95.2\\
\bottomrule
\multicolumn{8}{l}{\rule{0pt}{1em}\textsuperscript{1}ASE: Average Standard Error, \textsuperscript{2}ESE: Empirical Standard Error, \textsuperscript{3}RMSE: Root Mean Square Error}\\
\multicolumn{8}{l}{\rule{0pt}{1em}Prognostic variables $\rm  Z_1 \sim N(0,1)$, $\rm  Z_2 \sim Bin(0.5)$; Other variables $\rm  X_1 \sim N(0,1)$, $\rm  X_2 \sim Bin(0.5)$, $\rm  X_3 \sim U(0,2)$}\\
\end{tabular}
\vspace{1em}
\end{table}
\newpage
\textcolor{white}{Blablabla}

In practical applications, distinguishing prognostic variables $\rm  Z_j$ from non-prognostic ones $\rm  X_j$ can be challenging. There is a risk of model misspecification due to the accidental inclusion of variables that do not influence the survival outcome. Therefore, we also assessed the performance of the methods under model misspecification by including unrelated variables in the model in addition to the prognostic ones. We extended the analysis of RMST for scenarios 4 and 5 by adding one or multiple covariates, $\rm X_i$, unrelated to the survival as adjustment covariates. In both scenarios, the nuisances variables were simulated as $\rm  X_1 \sim N(0,1)$, $\rm  X_2 \sim Bin(0.5)$, and $\rm  X_3 \sim U(0, 2)$. The associated results for $n = 200$ correspond to the 3rd ($\rm X_1$) and 4th rows ($\rm X_1, X_2, X_3$) of each method for scenario 4 and the 4th row ($\rm X_1, X_2$) of each method for scenario 5 in Table \ref{orsini:tab2}. In scenario 4, no bias nor variance inflation was observed for all the methods. Similar results were observed for $\rm n= 500$ (supporting information table 8). For smaller sample sizes ($\rm n= 50, 100$), the impact of misspecification is more important: for the Bayesian GMM, the more unrelated variables were included, the larger the ASE became, while the frequentist approaches had only a slight increase of ASE with lower coverage rate than the nominal $95\%$ level (supporting information table 8). Similar results were observed with a delayed treatment effect (scenario 5): for $\rm n= 200$ (and $\rm n= 500$ supporting information Table 9), when adding unrelated variables, the estimations of the dRMST remained unbiased with ASE and ESE comparable to the other methods and satisfactory coverage rate. When the sample size decreases ($\rm n= 50$ and 100), the inflation of the ASE became more pronounced for the Bayesian GMM and resulted in large credible intervals that were conservative: $100\%$ coverage rate for $\rm n = 50$ when adjusting for $\rm  X_1$ and $\rm  X_2$ in addition to $\rm  Z_1$ and $\rm  Z_2$ (supporting information Table 9). However, the ESE of the Bayesian GMM did not inflate significantly, suggesting that the results may be strongly influenced by the wide prior we selected. Therefore, we extended the analysis in this particular setting (i.e. scenario 5, $\rm n= 50$ and adjusting the analysis for $\rm  Z_1, Z_2, X_1$ and $\rm  X_2$) by considering a normal prior with smaller variance, $\rm N(0,1)$, for all parameters, resulting in an ASE of $0.609$ compared to $2.091$ with the default prior of $\rm N(0,10)$, supporting this assumption. On the contrary, we did not observe an increase in ASE when adjusting for unrelated variables for the frequentist models. Consequently, for $\rm n = 50$, we observed confidence intervals narrower and resulting in a lower coverage rate than the nominal $95\%$ level: $92.2\%$ coverage rate for \citet{Zucker1998}, $91.9\%$ coverage rate for \citet{Andersen2004} (supporting information Table 9). 

\subsection{Model specification with an interaction term}

In scenario 6, the survival time was simulated according to a Weibull distribution with $\sigma = 0.8$ and $\rm \lambda = \exp(-1.2 + \log(HR_1)A_i + \log(HR_2)E_i + \log(HR_3)A_i \times E_i)$, where $\rm HR_1 = 1.7$, $\rm HR_2 = 0.5$, and $\rm HR_3 = 0.3$. Under these settings, the proportional hazards assumption is satisfied within each subgroup $\rm E = E^-$ and $\rm E = E^+$ but is clearly violated when considering the whole population with crossing survival curves (supporting information Figure 2). Firstly, we analyzed the simulated data of scenario 6 without taking into account biomarker $\rm E$ and its interaction with the treatment. When $\rm n\geq 100$, Bayesian GMM gave dRMST estimates between the two treatment arms comparable to the four benchmark methods with no bias and an appropriate coverage rate (supporting information Table 10). As observed in the previous scenarios, for $\rm n = 50$, the Bayesian GMM gave an ASE ($0.523$) slightly higher than the other methods (range from $0.494$ to $0.505$), leading to a coverage rate closer to the nominal $95\%$ level: $94.1\%$ for the Bayesian GMM and ranging from $92.0\%$ \citep{Zhang2023} to $92.7\%$ \citep{Zucker1998}. Secondly, we analyzed simulated data of scenario 6 by specifying a model including biomarker $\rm E$ and interaction term with treatment $\rm A\times E$. We limited this second analysis to pseudo-observation-based methods, as they are the only methods among those considered that allow for the inclusion of an interaction term in the model definition: the naive KM method and \citet{Zhang2023} approach do not allow for covariate adjustments, and \citet{Zucker1998} fit a Cox regression model stratified on the treatment variable to estimate the survival function conditional on the other covariates, before estimating an average survival function, averaging out the covariate parameters. Here, we consider the following model on RMST
\begin{equation}\label{interactionmodel1}
\rm \mu_i = E(\min(\tilde{T}_i, \tau)| A_i, E_i) = g^{-1}(\beta_0 + \delta A_i + \beta_1 E_i + \beta_2 A_i\times E_i).
\end{equation} The regression model (\ref{interactionmodel1}) can be expressed equivalently as
\begin{equation}\label{interactionmodel2}
\rm \mu_i = g^{-1}(\beta_0 + \beta_1 E_i + \delta^{-}A_i \mathbb{1}_{E_i = E^-} + \delta^{+}A_i \mathbb{1}_{E_i = E^+}),
\end{equation}
\noindent where $\rm \mathbb{1}_{E = E^-}$ and $\rm \mathbb{1}_{E = E^+}$ are indicator functions representing whether the biomarker for the $\rm i$-th individual was $\rm E^-$ or $\rm E^+$, respectively. This parametrization is adequate to estimate both parameters $\delta^- $, which corresponds to the dRMST between the two treatment arms for patients with $\rm E = E^-$, and $\delta^+ $, which corresponds to the dRMST between the two treatment arms for patients with $\rm E = E^+$. The parameter $\beta_1$ corresponds to the dRMST between the control patients with $\rm E^+$ and the control patients with $\rm E^-$. For $\rm n = 200$, we observed that both models produced valid estimates, ASE close to the ESE, and adequate coverage rate (Table \ref{orsini:tab3}).
\nobreak We note a slightly higher bias with the Bayesian GMM than in \citet{Andersen2004} for some coefficients. Still, the corresponding ESE was always smaller with the Bayesian GMM, resulting in an RMSE of the Bayesian GMM always smaller or equal to \citet{Andersen2004}. For higher sample size ($\rm n= 500$), the frequentist and Bayesian methods on pseudo-observations appeared equivalent in terms of bias, variance, and coverage. For smaller sample sizes (especially $\rm n = 50$), we observed a coverage slightly lower than the nominal $95\%$ level for \citet{Andersen2004} and slightly higher for the Bayesian GMM (supporting information Table 11).

\begin{table}[!ht]\centering
\caption{\label{orsini:tab3}Performance of the frequentist and Bayesian methods, based on pseudo-observations, on the estimation of the difference of restricted mean survival time (dRMST) between two treatment groups, adjusted on a binary covariate $\rm E$ with a predictive value on the treatment effect and an interaction term $\rm A \times E$, with non-proportional hazards treatment effect (scenario 6: crossing hazards), with $30\%$ of censoring and a sample size of $200$. The parameter $\delta^- = - 0.9025$ and $\delta^+ = 0.6492$ corresponds to the dRMST between the two treatment arms for patients with $\rm E = E^-$ and $\rm E = E^+$, respectively and $\beta_1 = 1.0644$ corresponds to the dRMST between $\rm E^+$ and $\rm E^-$ in control patients.}
\medskip
\begin{tabular}{cllrrrrr}
\toprule
Scenario & Methods & Estimated coefficient & Bias & ASE$^1$ & ESE$^2$ & RMSE$^3$ & Coverage\\
\midrule
\addlinespace[0.3em]
6 &\multicolumn{5}{l}{\textbf{Frequentist}}\\
&\hspace{1em}\citet{Andersen2004} & $\beta_1$  & 0.0074 & 0.330 & 0.340 & 0.340 & 93.9\\
&\hspace{1em}\citet{Andersen2004} & $\delta^-$ & 0.0060 & 0.307 & 0.310 & 0.310 & 94.0\\
&\hspace{1em}\citet{Andersen2004} & $\delta^+$ & 0.0021 & 0.292 & 0.309 & 0.309 & 92.4\\
\addlinespace[0.3em]
&\multicolumn{7}{l}{\textbf{Bayesian}}\\
&\hspace{1em}GMM & $\beta_1$ & 0.0212 & 0.339 & 0.333 & 0.334 & 95.2\\
&\hspace{1em}GMM & $\delta^-$ & 0.0258 & 0.315 & 0.305 & 0.306 & 94.6\\
&\hspace{1em}GMM & $\delta^+$ & 0.0022 & 0.301 & 0.306 & 0.306 & 93.8\\
\bottomrule
\multicolumn{8}{l}{\rule{0pt}{1em}\textsuperscript{1}ASE: Average Standard Error, \textsuperscript{2}ESE: Empirical Standard Error, \textsuperscript{3}RMSE: Root Mean Square Error}\\
\end{tabular}
\vspace{1em}
\end{table} 

\newpage
\section{Illustration on a real-data example }
\label{s:application}
\begin{figure}[!ht]\centering
\includegraphics[width=16.5cm]{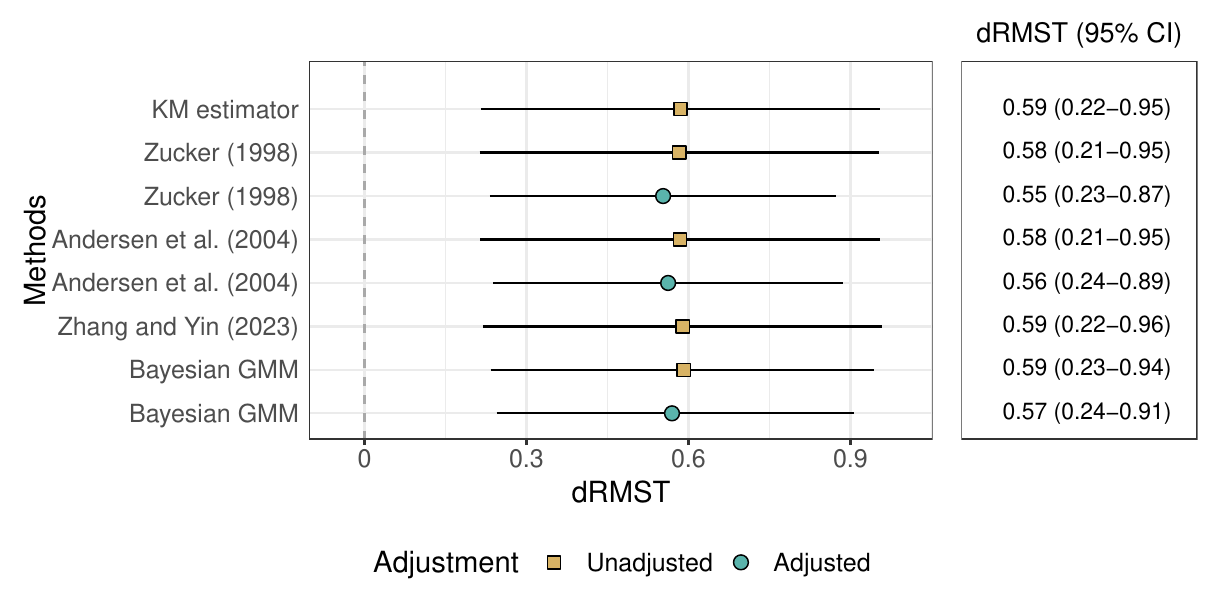}
\caption{Estimation of the difference of $5$-RMST between the ADT plus docetaxel and ADT alone arms for the Prostate-Specific Antigen (PSA) progression-free survival endpoint from the Getug-AFU 15 trial. The horizontal lines represent the $95\%$ confidence or credibility intervals (CI). The dashed vertical line represents no treatment effect.}
\label{orsini:fig2}
\end{figure}

We analyzed the data from the Getug-AFU 15, a randomized phase 3 trial comparing an androgen-deprivation therapy (ADT) alone ($\rm n=193$) or with docetaxel ($\rm n=192$) in non-castrate metastatic prostate cancer \citep{Gravis2013}. The median follow-up time was $4.2$ years. Our endpoint of interest was the Prostate-Specific Antigen (PSA) progression-free survival for which the PH assumption was rejected ($\rm p=0.00022$, \citet{Grambsch1994} test). We analyzed the complete case ($\rm n=345$ excluding $40$ patients from RCT due to missing covariates) to compare the RMST estimates.  $24\%$ of the patients were censored (supporting information Figure 3 shows the Kaplan-Meier curves for the two arms). The same methods, priors, and tuning parameters used for the simulations study were applied to the Getug-AFU 15 analysis. Figure \ref{orsini:fig2} reports the 5-year dRMST for the Bayesian GMM and the four benchmark methods with similar results for unadjusted dRMST estimates between the two arms. We observed an increase in precision of the dRMST estimates with narrower $95\%$ confidence or credibility intervals) adjusted for four binary variables: the Gleason score ($\geq 8$ vs. $<8$), European Cooperative Oncology Group performance status ($1-2$ vs. $0$), concentration of alkaline phosphatases (abnormal vs. normal), and presence of bone metastases (yes vs. no) with all methods allowing for covariates adjustment. These results are consistent with the findings of the simulation study. For Bayesian GMM, the trace plots of the NUTS sampling of the $3$ chains mixed well and appeared stationary, suggesting no divergence issue, supporting information Figure 4. 

\begin{figure}[!ht]\centering
\includegraphics[width=16.5cm]{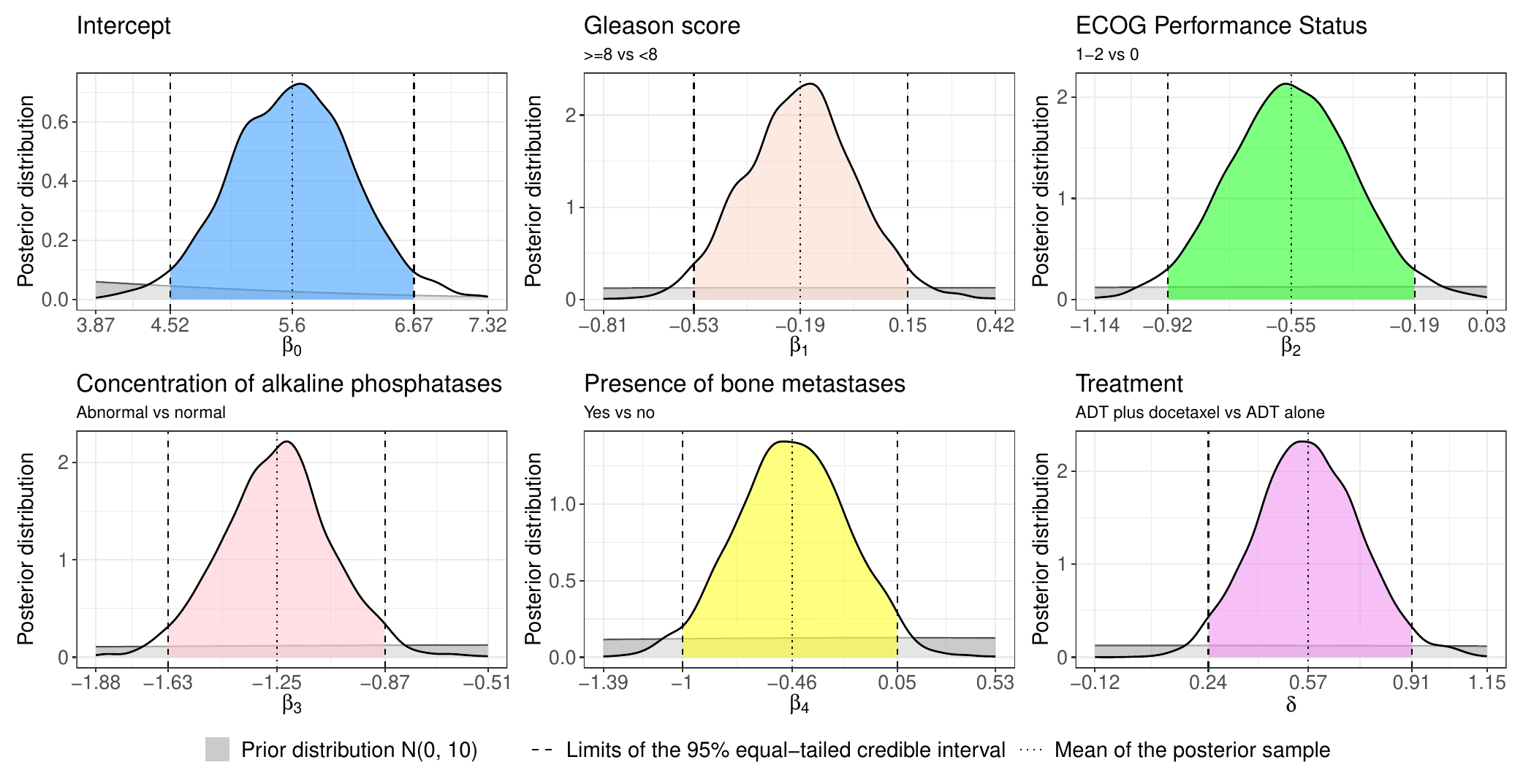}
\caption{Posterior distributions of parameters estimated from the Bayesian generalized method of moments applied to the Getug-AFU 15 trial for the PSA progression-free survival analysis. The intercept coefficient represents the 5-year RMST for the control group, with all binary covariates set to their reference levels. The remaining coefficients represent the 5-year dRMST between the two groups defined by each corresponding binary explanatory variable in the model.}
\label{orsini:fig3}
\end{figure}

With the Bayesian GMM, the adjusted $5$-dRMST was estimated to be $0.57$ year $(\rm 95\%\ CI:\ 0.24-0.91)$, meaning that receiving docetaxel in addition to ADT increases the lifetime without PSA progression over the next $5$ years by $0.57$ year, compared to receiving ADT alone. See supporting information Table 12 for the details of the parameter estimates with their standard deviation as well as the quantiles of the posterior distributions of the four covariates and the treatment indicator. One advantage of the pseudo-observations over the other approaches is the ability to estimate the baseline RMST, which corresponds to the intercept of the model. In the control group, with all explanatory binary covariates set to their reference levels, the estimated average RMST for an individual was $\beta_0 = 5.60$ years (SE $= 0.54$). All the other parameters of the model can be interpreted as $5$-dRMST for the treatment indicator but also any other explanatory variables. For example, $\beta_3$ can be interpreted as $5$-dRMST between patients with abnormal and normal concentrations of alkaline phosphatases. Moreover, the advantage of the Bayesian model we developed is to provide a tail posterior probability for any clinically meaningful threshold. For example, the estimated posterior probability of the dRMST between the two treatment arms over $5$ years to be higher than $3$ months: $\rm{P}(\delta \geq 3 \text{ months }) = 0.973$ with the Bayesian GMM adjusted for covariates with a Monte Carlo standard error (MC-SE) of $0.003$. For comparison, this posterior probability was estimated to be 0.963 (MC-SE $=0.004$) for the nonparametric model of \citet{Zhang2023}. As Bayesian GMM allows to take into account covariate adjustment, tail posterior probabilities may also be estimated for each covariate included in the model (Figure \ref{orsini:fig3}). For example, $\rm{P}(\beta_3 \leq - 1 \text{ year }) = 0.907$ (MC-SE $=0.005$), meaning that, on average, over the next five years, there is a high probability that patients with an abnormal concentration of alkaline phosphatases will experience PSA progression at least one year before the patients with a normal concentration of alkaline phosphatases. 

While the previous results reported the estimates of the dRMST for $\tau=5$ years, we conducted a sensitivity analysis to assess how varying the restricted time $\tau$ affects these results. We limited this sensitivity analysis to the unadjusted estimation of dRMST. Specifically, we examined $4$ alternatives for the choice of $\tau$ based on data-driven definitions: (a) the $90$-th percentile of the observed times, (b) the largest time for which the standard error of the survival estimate is lower a reasonable limit, either $5\%$ (b1) or $7.5\%$ (b2), as outlined by \citet{Karrison1997}, (c) the minimum of the maximum observed times in each treatment group, based on the approach suggested in \citet{Uno2022}. When applied to the Getug-AFU 15 trial, the alternative $\tau$ values ranged from $4.60$ years (a) to $6.66$ years (c), close to the original threshold of $5$ years, leading to results generally consistent with those presented in Figure \ref{orsini:fig2} (supporting information Table 13). The larger $\tau$ was, the larger the dRMST estimations were, with one exception for (c). This tendency was expected because we were integrating over a larger period of time with an experimental curve above the control curve until 6 years, see supporting information Figure 5. With large values of $\tau$, the proportion of subjects still at risk is small, resulting in larger confidence and credibility intervals.

\section{Discussion}
\label{s:discuss}
This paper presents a novel Bayesian approach for analyzing RMST with covariate adjustment using pseudo-observations in combination with the Bayesian GMM. The approach only requires the specification of a linear model but a model for the survival function is not needed, making it more straightforward than existing Bayesian methods. In addition, next to the posterior distribution of the treatment effect on RMST, posterior distributions of other covariate effects on the RMST can also be estimated, which is unique to this method.

While situations of non-PH are becoming increasingly common to analyze \citep{Lin2020} and alternative measures of the treatment effect are needed \citep{Uno2014}, this approach provides the option to adjust the dRMST estimation for covariates and interaction terms between the treatment and covariates in a Bayesian framework. Recently, oncologists are increasingly interested in Bayesian methods for designing RCTs \citep*{Goligher2024}, one attractive feature being the interesting interpretation of the results through the estimation of the whole posterior distribution of the parameters and not only point estimate with confidence bounds. As shown in the example of the Getug-AFU study, the Bayesian GMM enables the estimation of the probability of the dRMST between treatment groups (or any other covariate of interest) to exceed a desired threshold from the posterior distribution, facilitating meaningful decision-making. 

Simulation results indicate that the Bayesian GMM with pseudo-observations yielded results comparable to benchmark estimators for moderate ($\rm n=100,$ $200$) and large ($\rm n=500$) sample sizes in scenarios where survival is not related to covariates (scenario 1, 2, 3). When survival is related to one or two covariates, similar findings were observed between Bayesian GMM and other benchmark methods (scenarios 4 and 5). Omitting important variables produced less precision, and including covariates unrelated to the survival did not affect the treatment effect estimation for large samples $(\rm n\geq200)$. With a small sample size, the prior significantly impacted the posterior distribution. In such cases, using a wide prior may lead to excessive variability in parameter estimates, as the Bayesian algorithm is more likely to sample across a broad range of values. This issue became more pronounced as the number of parameters increased. In contrast, frequentist approaches \citep{Zucker1998, Andersen2004} have confidence intervals too narrow, potentially due to unverified asymptotic properties in small samples. Unlike the frequentist approaches, the inference of Bayesian methods only relies on the MCMC algorithm or its derivative, making them not prone to under-coverage for small sample sizes. The asymptotic proprieties of pseudo-observations, proved by \citet{Overgaard2017} and mentioned in section \ref{ss:RegressionPO}, as well as equation (\ref{Khisquarelimit}) are provided in this article to justify the Bayesian GMM in this context but do not play a role in the inference process. Nevertheless, it is rare to adjust for several variables with such a small sample size ($\rm n = 50$ and $30\%$ censoring), making this situation less common in real-world applications. One limit of our simulation study is to use a fixed $30\%$ censoring to focus more on simulating various scenarios with non-PH. However, such high censoring is rare in practical applications with a small sample size. We can extrapolate that the results will be consistent when considering lower censoring. The second limit is to have simulated only linear relationships of the covariates in the Weibull model. Caution must be taken regarding the potential misspecification of the model with continuous covariates since the functional form of the relationship between the covariate and the outcome must be specified.

When estimating dRMST, the choice of $\tau$ is important. In our simulation study, we considered the estimation of the difference of RMST at $\tau=5$ years. The sensitivity analysis on $\tau$ from the Getug-AFU 15 trial did not show an influence on the results, however in other settings, it may impact the RCT conclusion \citep{deBoissieu2024}. The ideal choice is to specify $\tau$ based on clinical considerations at the study design stage. Intuitively, one wants to choose the restricted time as the largest time point available to capture the maximum information from the Kaplan-Meier curve. However, in this application, it drastically increased the variance of the estimates due to the small number of patients still at risk. Further research is needed to generalize these findings to other settings. An approach to mitigate this arbitrary choice would be to estimate the dRMST at multiple time points. \citet*{Ambrogi2022} developed RMST curve estimations based on pseudo-observations. Based on this work, further research will extend the Bayesian GMM based on pseudo-observations to jointly analyze RMST at multiple time points. This can be achieved by considering not only one pseudo-observation per individual but a vector of pseudo-observations computed at different times and using a non-independence working correlation matrix. More generally, the Bayesian GMM using the pseudo-observations approach, which currently uses an independence working correlation matrix in this paper because patients are independent, can be extended to the RMST Bayesian analysis of correlated data using an appropriate working correlation structure as proposed by \cite{Vilain2023} for the frequentist analysis of RMST in a cluster randomized trial using an exchangeable working correlation structure.

In conclusion, the proposed Bayesian approach for RMST analysis with covariate adjustment provides straightforward, accurate results and meaningful interpretations, but careful consideration of sample size, prior choice, and choice of the restricted time is essential. 
%
\section*{Additional information}
The R code to analyze RMST with the Bayesian GMM on pseudo-observations is available on the Oncostat team's GitHub https://github.com/Oncostat/pseudo\_gmm

\section*{Acknowledgments}
The authors thank C. Zhang for providing the code to implement his Bayesian nonparametric model and for his time and dedication in answering their questions. 

This work was awarded the PhD Student Award of the QuanTIM Webinars organized by the SESSTIM research unit (https://sesstim.univ-amu.fr/phd-student-award).

 \section*{Conflict of interest}

 The authors declare that they have no conflict of interest.

 \section*{Funding}
 This study was funded by PhD grant MESRI from the doctoral School of Public Health, Paris-Saclay University. 



\end{document}


\maketitle
 \begin{figure*}[!ht]\centering
\includegraphics[width=16.5cm]{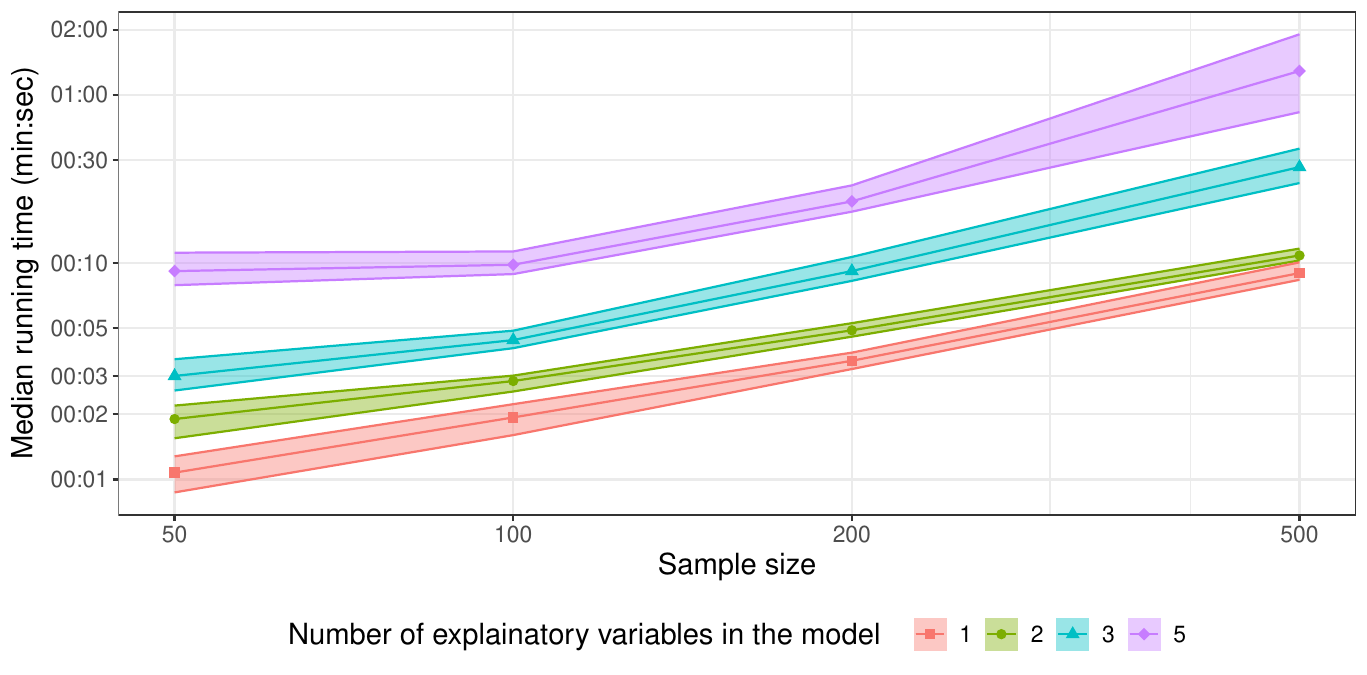}
\caption{Running times of one chain for the Bayesian GMM using pseudo-observations in scenario 5 (delayed treatment effect) according to sample size and number of explanatory variables included in the model. The x and y axes are on the log scale for a clearer visualization. The colored areas represent the interquartile range.}
\label{orsini:fig2}
\end{figure*}

 \begin{figure*}[!ht]\centering
\includegraphics[scale = 0.75]{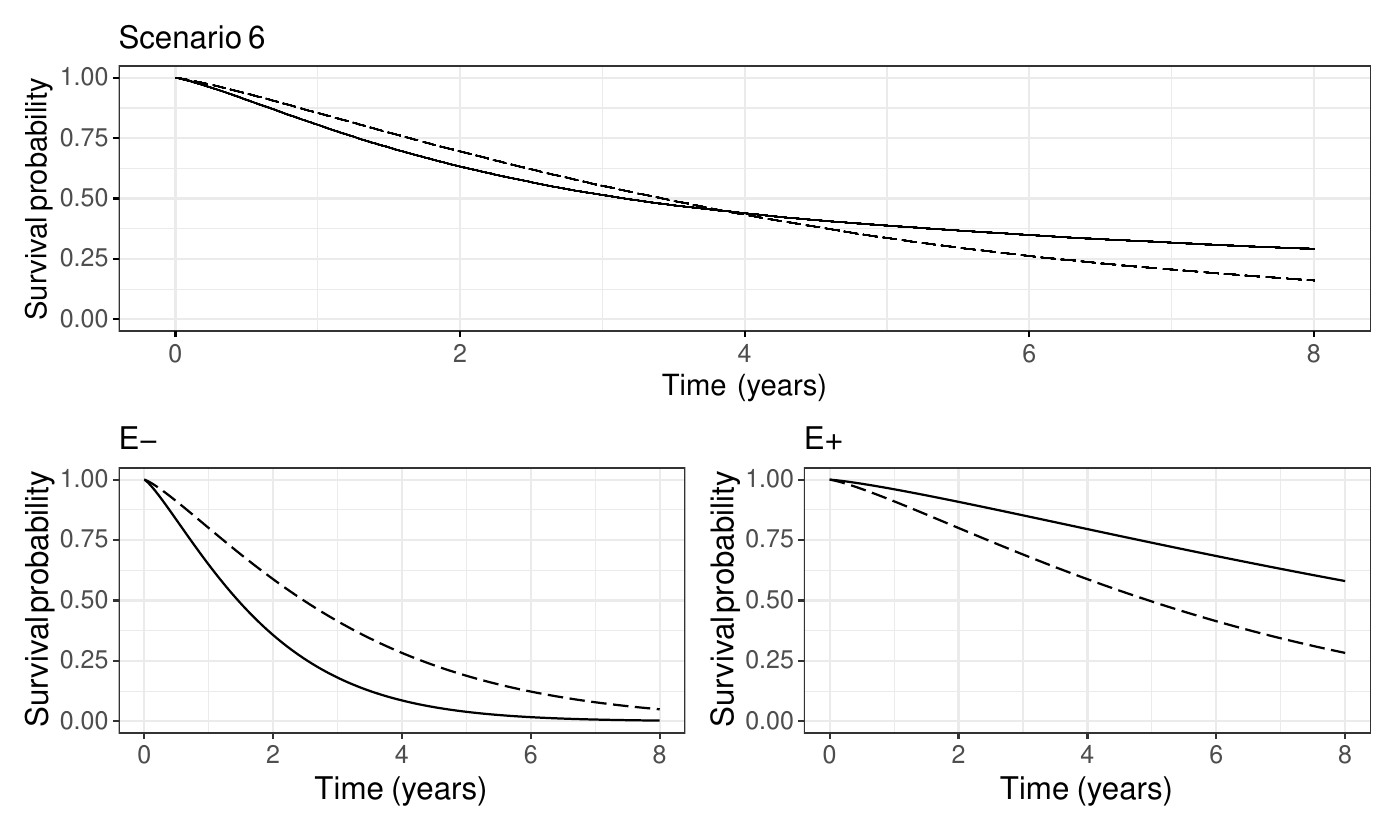}
\caption{True survival curves of scenario 6, in each subgroup of the biomarker variable, $\rm{E=(E^-,E^+)}$ (bottom panel) and on the whole population (top panel). These simulation settings produce a case where the proportional hazards assumption is verified within each subgroup of the same sample size $E = E^-$ and $E = E^+$ but produce crossing survival curves when considering the whole population.}
\label{orsini:fig1}
\end{figure*}

 \begin{figure*}[!ht]\centering
\includegraphics[width=16.5cm]{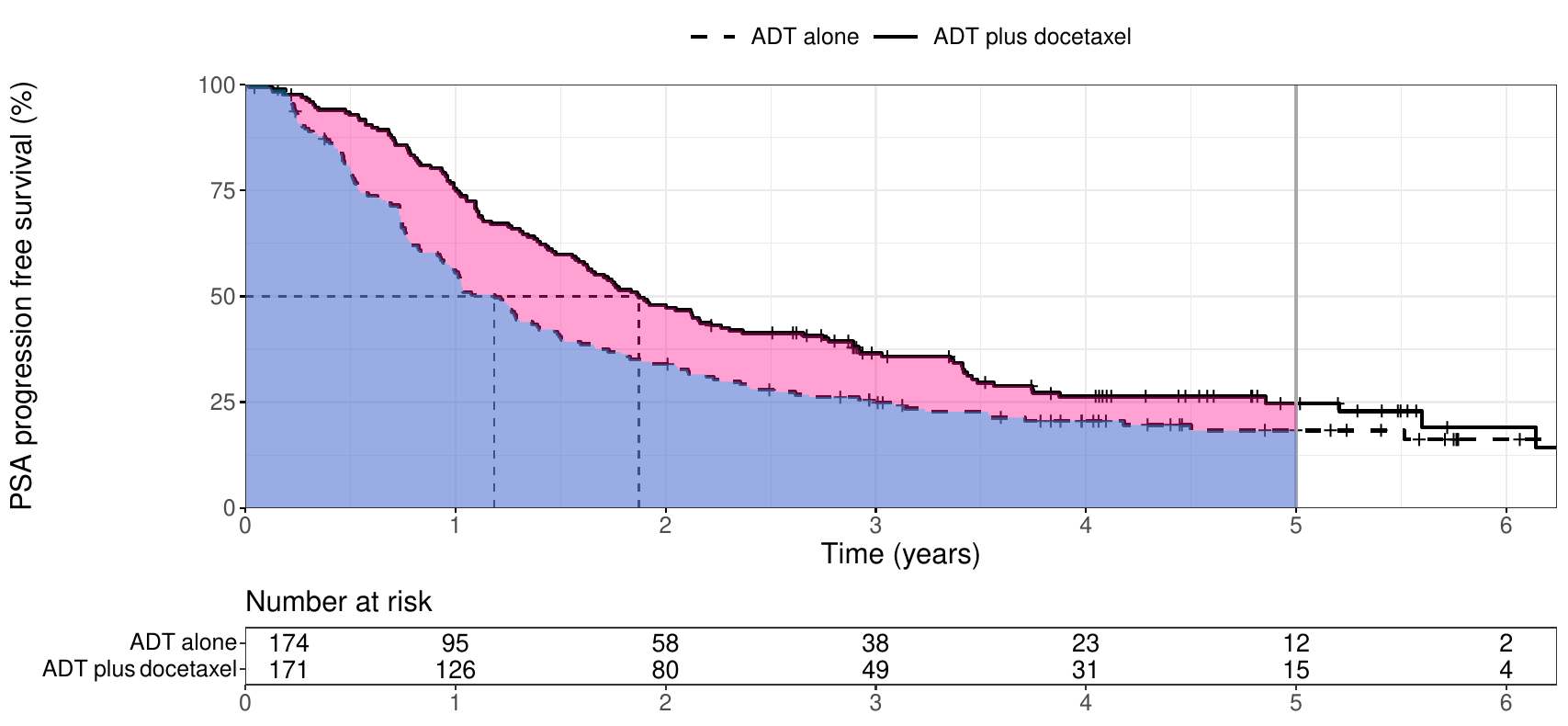}
\caption{Kaplan-Meier curves for the Prostate-Specific Antigen (PSA) progression-free survival from the Getug-AFU 15 trial. The blue area under the survival curve until $\tau = 5$ year (vertical grey line) represents the $5$-RMST of the control group (ADT alone) and the pink area represents the 5 years- difference of restricted mean survival time (dRMST) between the two arms.}
\label{orsini:fig3}
\end{figure*}

 \begin{figure*}[!ht]\centering
\includegraphics[width=16.5cm]{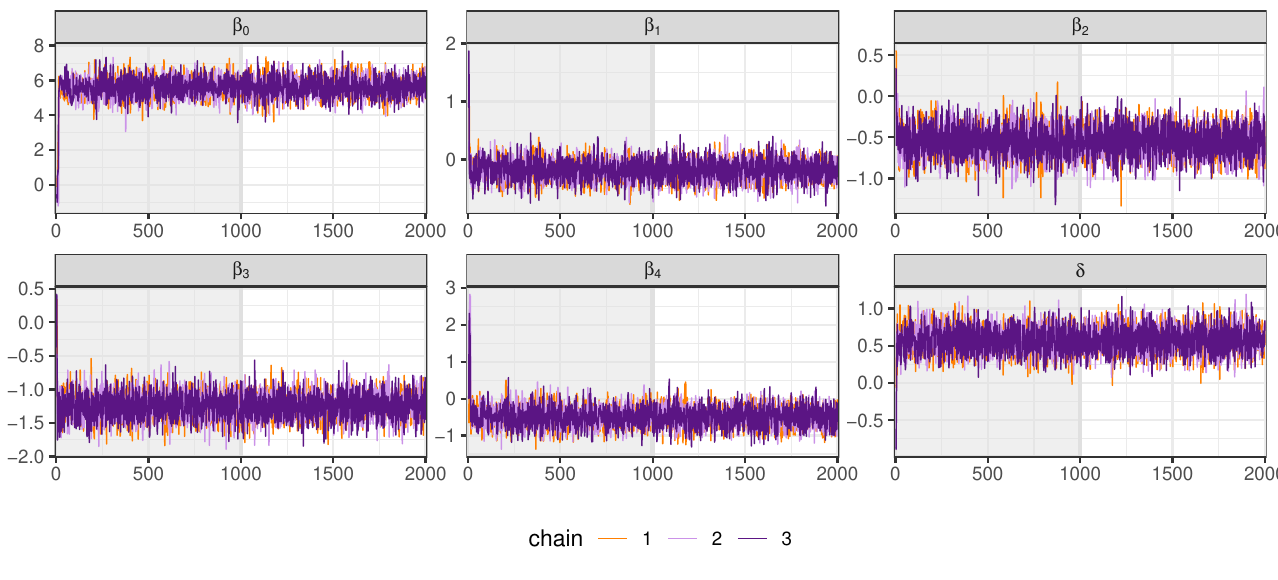}
\caption{Trace plots of the three Markov chains for each parameter of the Bayesian GMM model ($\beta_0$: intercept, $\beta_1$: Gleason score, $\beta_2$: European Cooperative Oncology Group performance, $\beta_3$: concentration of alkaline phosphatases, $\beta_4$: presence of bone metastases and $\delta$: treatment effect) in the post hoc analysis of the Getug-AFU 15 trial. The grey area represents the burn-in period.}
\label{orsini:fig4}
\end{figure*}

 \begin{figure*}[!ht]\centering
\includegraphics[width = 16.5cm]{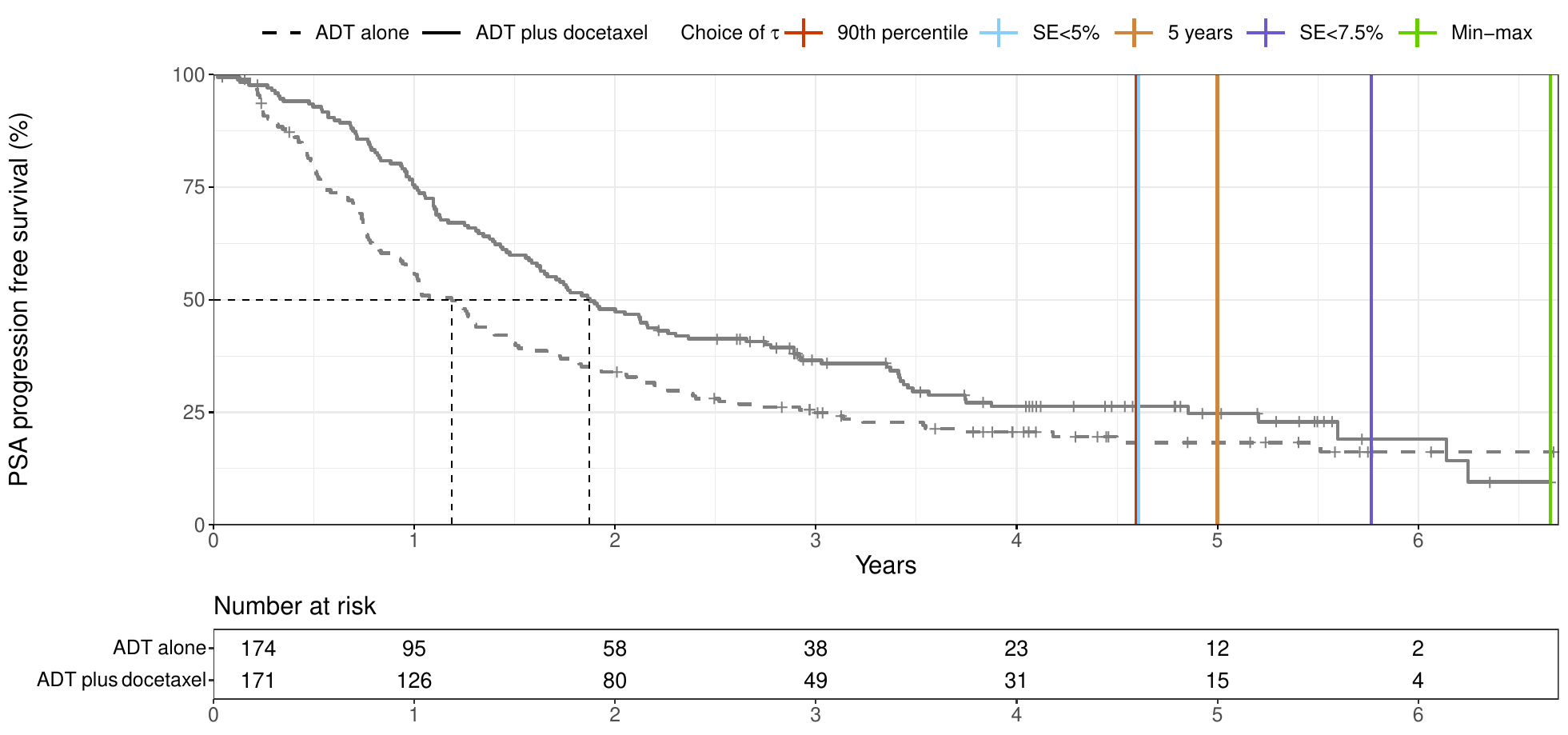}
\caption{Kaplan-Meier curves for the PSA progression-free survival from the Getug-AFU 15 trial. The vertical orange line represents the restricted time $\tau = 5$ year. The other vertical lines represent alternative definitions of $\tau$: (a) the $90$-th percentile of the observed times (brown), (b) the largest time point $\rm t$ for which the standard error of the survival estimate is lower than a reasonable limit, either (b1) $5\%$ (light blue) or (b2) $7.5\%$ (dark blue), (c) the minimum of the maximum observed times in each treatment group (green).}
\label{orsini:fig5}
\end{figure*}

\begin{sidewaystable}[!ht]
\caption{Details of the different simulated scenarios}
\label{orsini:tab1}
\centering
    \scalebox{0.93}{
\begin{tabular}{llll}
\toprule
Scenario & True model & Shape $\sigma$ & Scale $\lambda$ \\ 
\midrule
\addlinespace[0.3em]
1 & $\rm S(t) = \exp(-(\lambda t)^{1/\sigma})$ & $\sigma = 0.8$ & $\lambda= \exp(-1.2 + \log(0.6)\rm{A})$ \\
\midrule
\addlinespace[0.3em]
2 & $\rm S(t) = \exp(-(\lambda t)^{1/\sigma})$ & $\sigma = 1.33\times(1-\rm{A}) + 0.67\times \rm{A}$
 & $\lambda= 0.2\times (1-\rm{A}) + 0.18\times\rm{A}$ \\ 
\midrule 
\addlinespace[0.3em]
3 & $\rm S(t) = \exp(-(\lambda t)^{1/\sigma})$ & $\sigma = 0.6\times(1-\rm{A}) + 0.8 \times \rm{A}$ & $\lambda= 0.28\times (1-\rm{A}) + 0.18\times\rm{A}$ \\ 
\midrule 
\addlinespace[0.3em]
4 & $\rm S(t) = \exp(-(\lambda t)^{1/\sigma}\exp(\beta_1 \rm{\rm{Z_1}}))$ & $\sigma = 1.33\times (1-\rm{A}) + 0.67\times\rm{A}$ & $\lambda= 0.2\times (1-\rm{A}) + 0.18\times \rm{A}$ \\ 
\midrule 
\addlinespace[0.3em]
5 & $\rm S(t) = \exp(-(\lambda t)^{1/\sigma}\exp(\beta_1 \rm{\rm{Z_1}} + \beta_2\rm{Z_2}))$ & $\sigma = 0.6\times(1-\rm{A}) + 0.8\times \rm{A}$ & $\lambda= 0.28\times (1-\rm{A}) + 0.18\times\rm{A}$ \\ 
\midrule 
\addlinespace[0.3em]
6 & $\rm S(t) = \exp(-(\lambda t)^{1/\sigma})$ & $\sigma = 0.8$ & $\lambda= \exp(-1.2 + \log(1.7)\rm{A} + \log(0.5)\rm{E} + \log(0.3)\rm{A} \times \rm{E})$ \\ 
\bottomrule
\multicolumn{4}{l}{\rule{0pt}{1em}$\rm{A}$ a binary treatment indicator ($\rm{A} = 1$ for experimental arm and $\rm{A} = 0$ for control arm) with $\rm{A} \sim Bernoulli(0.5)$}\\
\multicolumn{4}{l}{\rule{0pt}{1em}$\rm{E}$ is a binary biomarker with a predictive value on the treatment effect ($\rm{E} = \rm{E^+} = 1$ and $\rm{E} = \rm{E^-} = 0 $) with $\rm{E} \sim Bernoulli(0.5)$}\\
\multicolumn{4}{l}{\rule{0pt}{1em}Prognostic variables: $\rm{Z_1} \sim U(0,2)$ (scenario 4), $\rm{Z_1} \sim N(0,1)$ and $\rm{Z_2} \sim Bin(0.5)$ (scenario 5)}\\
\multicolumn{4}{l}{\rule{0pt}{1em} Other variables unrelated to the survival function were simulated to evaluate model misspecification in scenarios 4 and 5:}\\ 
\multicolumn{4}{l}{\rule{0pt}{1em} $\rm{X_1} \sim N(0,1)$, $\rm{X_2} \sim Bin(0.5)$, $\rm{X_3} \sim U(0,2)$} \\
\end{tabular}
}
\end{sidewaystable}

\begin{table}[h!]
    \centering
        \caption{Percentage of replications where the last simulated observed time in both arms of the randomized controlled trial is less than the restricted time $\tau = 5$, so $\tau$ was redefined as the minimum of the maximum observed times of each arm.}
    \label{tab:your_label_here}
    \begin{tabular}{lrrrr}
        \toprule
        & $\rm n = 50$ & $\rm n = 100$ & $\rm n = 200$ & $\rm n = 500$ \\
        \midrule
        Scenario 1 & 5.2\% & 0.1\% & 0\% & 0\% \\
        Scenario 2 & 3.6\% & 0\% & 0\% & 0\% \\
        Scenario 3 & 5.3\% & 0.2\% & 0\% & 0\% \\
        Scenario 4 & 24.2\% & 3.0\% & 0.1\% & 0\% \\
        Scenario 5 & 9.4\% & 1.1\% & 0.4\% & 0.1\% \\
        Scenario 6 & 0.2\% & 0\% & 0\% & 0\% \\
        \bottomrule
    \end{tabular}
\end{table}

\begin{table}[h!]
    \centering
        \caption{Percentage of replications with non convergence (i.e. with $\widehat{R}\geq1.1$) for the Bayesian GMM per scenario according to different model specifications and sample sizes. Note that these replications are removed from the analysis.}
    \label{tab:your_label_here2}
    \begin{tabular}{llrrrr}
        \toprule
        Scenario & Adjustment Variables & $\rm n = 50$ & $\rm n = 100$ & $\rm n = 200$ & $\rm n = 500$ \\
        \midrule
    Scenario 1 & - & 0\% & 0\% & 0\% & 0\% \\
    \midrule
    Scenario 2 & - & 0\% & 0\% & 0\% & 0\% \\
    \midrule
    Scenario 3 & - & 0\% & 0\% & 0\% & 0\% \\
    \midrule
    Scenario 4 & - & 0\% & 0\% & 0\% & 0\% \\
    & $\rm{Z_1}$ & 0\% & 0\% & 0\% & 0\% \\
    & $\rm{Z_1}$, $\rm{X_1}$ & 0.5\% & 0.2\% & 0\% & 0\% \\
    & $\rm{Z_1}$, $\rm{X_1}$, $\rm{X_2}$, $\rm{X_3}$ & 1.3\% & 0.7\% & 0\% & 0\% \\
    \midrule
    Scenario 5 & - & 0\% & 0\% & 0\% & 0\% \\
    & $\rm{Z_1}$ & 0.2\% & 0.1\% & 0\% & 0.1\% \\
    & $\rm{Z_1}$, $\rm{Z_2}$ & 0.6\% & 0.1\% & 0.1\% & 0.1\% \\
    & $\rm{Z_1}$, $\rm{Z_2}$, $\rm{X_1}$, $\rm{X_2}$ & 1.2\% & 1.5\% & 0.4\% & 0.1\%  \\
    \midrule
    Scenario 6 & - & 0\% & 0\% & 0\% & 0\% \\
    & $\rm E$, $\rm E \times A$ & 0\% & 0\% & 0\% & 0\% \\
        \bottomrule
        \multicolumn{6}{l}{\rule{0pt}{1em}Prognostic variables $\rm{Z_1} \sim N(0,1)$ and $\rm{Z_2} \sim Bin(0.5)$}\\
        \multicolumn{6}{l}{\rule{0pt}{1em}Other variables unrelated to the survival function $\rm{X_1} \sim N(0,1)$, $\rm{X_2} \sim Bin(0.5)$, and $\rm{X_3} \sim U(0,2)$}\\
        \multicolumn{6}{l}{\rule{0pt}{1em}$\rm{E} \sim Bin(0.5)$}\\
    \end{tabular}
\end{table}

\begin{table}[!ht]\centering

\caption{\label{orsini:tab2}Performance of the frequentist and Bayesian methods on the univariate estimation of the difference of restricted mean survival time (dRMST) between two treatment groups under the proportional hazards assumption (scenario 1: null hypothesis true dRMST, $\delta = 0$), with $30\%$ of censoring and different sample sizes.}
\medskip
\begin{tabular}{llrrrrr}
\toprule
n & Methods & Bias & ASE$^1$& ESE$^2$  & RMSE$^3$ & Coverage\\
\midrule
\addlinespace[0.3em]
50&\multicolumn{5}{l}{\textbf{Frequentist}}\\
&\hspace{1em}KM estimator & -0.0163 & 0.490 & 0.510 & 0.510 & 93.1\\
&\hspace{1em}Zucker (1998) & -0.0138 & 0.499 & 0.510 & 0.511 & 94.0\\
&\hspace{1em}Andersen et al. (2004) & -0.0166 & 0.494 & 0.510 & 0.511 & 93.3\\
\addlinespace[0.3em]
&\multicolumn{6}{l}{\textbf{Bayesian}}\\
&\hspace{1em}Zhang and Yin (2023) & -0.0159 & 0.482 & 0.528 & 0.528 & 92.9\\
&\hspace{1em}GMM & 0.0171 & 0.511 & 0.493 & 0.493 & 95.4\\
\midrule
\addlinespace[0.3em]
100&\multicolumn{5}{l}{\textbf{Frequentist}}\\
&\hspace{1em}KM estimator & 0.0000 & 0.354 & 0.360 & 0.360 & 93.8\\
&\hspace{1em}Zucker (1998) & 0.0003 & 0.355 & 0.356 & 0.356 & 94.0\\
&\hspace{1em}Andersen et al. (2004) & -0.0003 & 0.355 & 0.360 & 0.360 & 93.6\\
\addlinespace[0.3em]
&\multicolumn{6}{l}{\textbf{Bayesian}}\\
&\hspace{1em}Zhang and Yin (2023) & 0.0001 & 0.349 & 0.361 & 0.361 & 93.4\\
&\hspace{1em}GMM & 0.0152 & 0.359 & 0.354 & 0.354 & 94.2\\
\midrule
\addlinespace[0.3em]
200&\multicolumn{5}{l}{\textbf{Frequentist}}\\
&\hspace{1em}KM estimator & -0.0050 & 0.251 & 0.247 & 0.247 & 95.7\\
&\hspace{1em}Zucker (1998) & -0.0050 & 0.251 & 0.246 & 0.246 & 95.8\\
&\hspace{1em}Andersen et al. (2004) & -0.0049 & 0.251 & 0.247 & 0.247 & 95.7\\
\addlinespace[0.3em]
&\multicolumn{6}{l}{\textbf{Bayesian}}\\
&\hspace{1em}Zhang and Yin (2023) & -0.0049 & 0.249 & 0.247 & 0.247 & 95.2\\
&\hspace{1em}GMM & 0.0022 & 0.252 & 0.245 & 0.245 & 95.8\\
\midrule
\addlinespace[0.3em]
500&\multicolumn{5}{l}{\textbf{Frequentist}}\\
&\hspace{1em}KM estimator & 0.0039 & 0.159 & 0.158 & 0.158 & 95.1\\
&\hspace{1em}Zucker (1998) & 0.0039 & 0.159 & 0.157 & 0.157 & 95.1\\
&\hspace{1em}Andersen et al. (2004) & 0.0039 & 0.159 & 0.158 & 0.158 & 95.1\\
\addlinespace[0.3em]
&\multicolumn{6}{l}{\textbf{Bayesian}}\\
&\hspace{1em}Zhang and Yin (2023) & 0.0040 & 0.158 & 0.158 & 0.158 & 94.8\\
&\hspace{1em}GMM & 0.0061 & 0.159 & 0.157 & 0.157 & 95.0\\
\bottomrule
\multicolumn{6}{l}{\rule{0pt}{1em}\textsuperscript{1}ASE: Average Standard Error}\\
\multicolumn{6}{l}{\rule{0pt}{1em}\textsuperscript{2}ESE: Empirical Standard Error}\\
\multicolumn{6}{l}{\rule{0pt}{1em}\textsuperscript{3}RMSE: Root Mean Square Error}\\
\end{tabular}
\end{table}

\begin{table}[!ht]\centering

\caption{\label{orsini:tab3}Performance of the frequentist and Bayesian methods on the univariate estimation of the difference of restricted mean survival time (dRMST) between two treatment groups under the proportional hazards assumption (scenario 1: alternative hypothesis true dRMST, $\delta = 0.8196$), with $30\%$ of censoring and different sample sizes.}
\medskip
\begin{tabular}{llrrrrr}
\toprule
n & Methods & Bias & ASE$^1$& ESE$^2$  & RMSE$^3$ & Coverage\\
\midrule
\addlinespace[0.3em]
50&\multicolumn{5}{l}{\textbf{Frequentist}}\\
&\hspace{1em}KM estimator & -0.0214 & 0.481 & 0.512 & 0.513 & 92.4\\
&\hspace{1em}Zucker (1998) & -0.0332 & 0.480 & 0.506 & 0.507 & 92.8\\
&\hspace{1em}Andersen et al. (2004) & -0.0210 & 0.484 & 0.513 & 0.514 & 92.7\\
\addlinespace[0.3em]
& \multicolumn{5}{l}{\textbf{Bayesian}}\\
&\hspace{1em}Zhang and Yin (2023)  & -0.0138 & 0.470 & 0.520 & 0.520 & 92.1\\
&\hspace{1em}GMM & -0.0082 & 0.501 & 0.498 & 0.498 & 94.4\\
\midrule
\addlinespace[0.3em]
100&\multicolumn{5}{l}{\textbf{Frequentist}}\\
&\hspace{1em}KM estimator & 0.0061 & 0.344 & 0.353 & 0.353 & 93.7\\
&\hspace{1em}Zucker (1998) & -0.0037 & 0.343 & 0.349 & 0.349 & 94.1\\
&\hspace{1em}Andersen et al. (2004) & 0.0059 & 0.346 & 0.352 & 0.352 & 94.1\\
\addlinespace[0.3em]
&\multicolumn{6}{l}{\textbf{Bayesian}}\\
&\hspace{1em}Zhang and Yin (2023) & 0.0059 & 0.340 & 0.353 & 0.353 & 93.5\\
&\hspace{1em}GMM & 0.0106 & 0.350 & 0.348 & 0.348 & 94.5\\
\midrule
\addlinespace[0.3em]
200&\multicolumn{5}{l}{\textbf{Frequentist}}\\
&\hspace{1em}KM estimator & -0.0071 & 0.244 & 0.248 & 0.248 & 94.4\\
&\hspace{1em}Zucker (1998) & -0.0119 & 0.244 & 0.246 & 0.247 & 94.7\\
&\hspace{1em}Andersen et al. (2004) & -0.0069 & 0.245 & 0.248 & 0.248 & 94.3\\
\addlinespace[0.3em]
&\multicolumn{6}{l}{\textbf{Bayesian}}\\
&\hspace{1em}Zhang and Yin (2023) & -0.0068 & 0.243 & 0.248 & 0.248 & 94.6\\
&\hspace{1em}GMM & -0.0055 & 0.246 & 0.246 & 0.246 & 94.8\\
\midrule
\addlinespace[0.3em]
500&\multicolumn{5}{l}{\textbf{Frequentist}}\\
&\hspace{1em}KM estimator & 0.0060 & 0.154 & 0.155 & 0.155 & 94.7\\
&\hspace{1em}Zucker (1998) & 0.0040 & 0.154 & 0.155 & 0.155 & 94.9\\
&\hspace{1em}Andersen et al. (2004) & 0.0060 & 0.155 & 0.155 & 0.155 & 94.8\\
\addlinespace[0.3em]
&\multicolumn{6}{l}{\textbf{Bayesian}}\\
&\hspace{1em}Zhang and Yin (2023) & 0.0061 & 0.154 & 0.155 & 0.155 & 94.6\\
&\hspace{1em}GMM & 0.0061 & 0.155 & 0.155 & 0.155 & 95.0\\
\bottomrule
\multicolumn{6}{l}{\rule{0pt}{1em}\textsuperscript{1}ASE: Average Standard Error}\\
\multicolumn{7}{l}{\rule{0pt}{1em}\textsuperscript{2}ESE: Empirical Standard Error}\\
\multicolumn{6}{l}{\rule{0pt}{1em}\textsuperscript{3}RMSE: Root Mean Square Error}\\
\end{tabular}
\end{table}


\begin{table}[!ht]\centering
\caption{\label{orsini:tab4}Performance of the frequentist and Bayesian methods on the univariate estimation of the difference of restricted mean survival time (dRMST) between two treatment groups with non-proportional hazards treatment effect (scenario 2: early effect true dRMST, $\delta = 0.7302$), with $30\%$ of censoring and different sample sizes.}
\medskip
\begin{tabular}{llrrrrr}
\toprule
n & Methods & Bias & ASE$^1$ & ESE$^2$  & RMSE$^3$ & Coverage\\
\midrule
\addlinespace[0.3em]
50&\multicolumn{5}{l}{\textbf{Frequentist}}\\
&\hspace{1em}KM estimator & -0.0056 & 0.498 & 0.538 & 0.538 & 92.2\\
&\hspace{1em}Zucker (1998) & -0.0198 & 0.493 & 0.526 & 0.526 & 92.3\\
&\hspace{1em}Andersen et al. (2004) & -0.0058 & 0.499 & 0.538 & 0.538 & 92.1\\
\addlinespace[0.3em]
&\multicolumn{6}{l}{\textbf{Bayesian}}\\
&\hspace{1em}Zhang and Yin (2023) & -0.0061 & 0.487 & 0.538 & 0.538 & 92.1\\
&\hspace{1em}GMM & 0.0213 & 0.514 & 0.521 & 0.522 & 93.2\\
\midrule
\addlinespace[0.3em]
100&\multicolumn{5}{l}{\textbf{Frequentist}}\\
&\hspace{1em}KM estimator & 0.0068 & 0.356 & 0.356 & 0.356 & 94.3\\
&\hspace{1em}Zucker (1998) & -0.0006 & 0.354 & 0.352 & 0.352 & 94.6\\
&\hspace{1em}Andersen et al. (2004) & 0.0068 & 0.356 & 0.356 & 0.356 & 94.4\\
\addlinespace[0.3em]
&\multicolumn{6}{l}{\textbf{Bayesian}}\\
&\hspace{1em}Zhang and Yin (2023) & 0.0067 & 0.352 & 0.356 & 0.356 & 94.4\\
&\hspace{1em}GMM & 0.0192 & 0.359 & 0.351 & 0.351 & 94.8\\
\midrule
\addlinespace[0.3em]
200&\multicolumn{5}{l}{\textbf{Frequentist}}\\
&\hspace{1em}KM estimator & -0.0071 & 0.252 & 0.254 & 0.254 & 95.4\\
&\hspace{1em}Zucker (1998) & -0.0107 & 0.251 & 0.252 & 0.253 & 95.3\\
&\hspace{1em}Andersen et al. (2004) & -0.0070 & 0.252 & 0.254 & 0.254 & 95.4\\
\addlinespace[0.3em]
&\multicolumn{6}{l}{\textbf{Bayesian}}\\
&\hspace{1em}Zhang and Yin (2023) & -0.0072 & 0.251 & 0.254 & 0.254 & 95.3\\
&\hspace{1em}GMM & -0.0020 & 0.252 & 0.252 & 0.252 & 95.4\\
\midrule
\addlinespace[0.3em]
500&\multicolumn{5}{l}{\textbf{Frequentist}}\\
&\hspace{1em}KM estimator & 0.0063 & 0.160 & 0.160 & 0.160 & 95.0\\
&\hspace{1em}Zucker (1998) & 0.0048 & 0.160 & 0.160 & 0.160 & 95.2\\
&\hspace{1em}Andersen et al. (2004) & 0.0063 & 0.160 & 0.160 & 0.160 & 95.0\\\addlinespace[0.3em]
&\multicolumn{6}{l}{\textbf{Bayesian}}\\
&\hspace{1em}Zhang and Yin (2023) & 0.0065 & 0.159 & 0.160 & 0.161 & 95.1\\
&\hspace{1em}GMM & 0.0074 & 0.159 & 0.160 & 0.160 & 95.2\\
\bottomrule
\multicolumn{6}{l}{\rule{0pt}{1em}\textsuperscript{1}ASE: Average Standard Error}\\
\multicolumn{7}{l}{\rule{0pt}{1em}\textsuperscript{2}ESE: Empirical Standard Error}\\
\multicolumn{6}{l}{\rule{0pt}{1em}\textsuperscript{3}RMSE: Root Mean Square Error}\\
\end{tabular}
\end{table}

\begin{table}[!ht]\centering
\caption{\label{orsini:tab5}Performance of the frequentist and Bayesian methods on the univariate estimation of the difference of restricted mean survival time (dRMST) between two treatment groups with non-proportional hazards treatment effect (scenario 3: late effect true dRMST, $\delta = 0.5644$), with $30\%$ of censoring and different sample sizes.}
\medskip
\begin{tabular}{llrrrrr}
\toprule
n & Methods & Bias & ASE$^1$ & ESE$^2$  & RMSE$^3$ & Coverage\\
\midrule
\addlinespace[0.3em]
50&\multicolumn{5}{l}{\textbf{Frequentist}}\\
&\hspace{1em}KM estimator & -0.0212 & 0.461 & 0.491 & 0.492 & 92.7\\
&\hspace{1em}Zucker (1998) & -0.0278 & 0.459 & 0.485 & 0.486 & 93.1\\
&\hspace{1em}Andersen et al. (2004) & -0.0206 & 0.464 & 0.492 & 0.492 & 93.1\\\addlinespace[0.3em]
&\multicolumn{6}{l}{\textbf{Bayesian}}\\
&\hspace{1em}Zhang and Yin (2023) & -0.0129 & 0.451 & 0.497 & 0.497 & 92.3\\
&\hspace{1em}GMM & -0.0044 & 0.482 & 0.480 & 0.480 & 94.4\\
\midrule
\addlinespace[0.3em]
100&\multicolumn{5}{l}{\textbf{Frequentist}}\\
&\hspace{1em}KM estimator & 0.0058 & 0.330 & 0.340 & 0.340 & 93.5\\
&\hspace{1em}Zucker (1998) & -0.0016 & 0.329 & 0.336 & 0.336 & 93.9\\
&\hspace{1em}Andersen et al. (2004) & 0.0055 & 0.331 & 0.339 & 0.339 & 93.7\\
\addlinespace[0.3em]
&\multicolumn{6}{l}{\textbf{Bayesian}}\\
&\hspace{1em}Zhang and Yin (2023) & 0.0059 & 0.326 & 0.340 & 0.340 & 93.4\\
&\hspace{1em}GMM & 0.0124 & 0.336 & 0.335 & 0.335 & 94.3\\
\midrule
\addlinespace[0.3em]
200&\multicolumn{5}{l}{\textbf{Frequentist}}\\
&\hspace{1em}KM estimator & -0.0065 & 0.234 & 0.237 & 0.237 & 93.9\\
&\hspace{1em}Zucker (1998) & -0.0101 & 0.233 & 0.235 & 0.236 & 94.0\\
&\hspace{1em}Andersen et al. (2004) & -0.0063 & 0.235 & 0.237 & 0.237 & 93.9\\
\addlinespace[0.3em]
&\multicolumn{6}{l}{\textbf{Bayesian}}\\
&\hspace{1em}Zhang and Yin (2023) & -0.0065 & 0.232 & 0.237 & 0.237 & 94.1\\
&\hspace{1em}GMM & -0.0035 & 0.236 & 0.236 & 0.236 & 95.1\\
\midrule
\addlinespace[0.3em]
500&\multicolumn{5}{l}{\textbf{Frequentist}}\\
&\hspace{1em}KM estimator & 0.0052 & 0.148 & 0.149 & 0.149 & 94.7\\
&\hspace{1em}Zucker (1998) & 0.0036 & 0.148 & 0.149 & 0.149 & 94.9\\
&\hspace{1em}Andersen et al. (2004) & 0.0052 & 0.149 & 0.149 & 0.149 & 94.6\\
\addlinespace[0.3em]
&\multicolumn{6}{l}{\textbf{Bayesian}}\\
&\hspace{1em}Zhang and Yin (2023) & 0.0051 & 0.148 & 0.149 & 0.149 & 94.6\\
&\hspace{1em}GMM & 0.0060 & 0.149 & 0.149 & 0.149 & 94.8\\
\bottomrule
\multicolumn{7}{l}{\rule{0pt}{1em}\textsuperscript{1}ASE: Average Standard Error}\\
\multicolumn{7}{l}{\rule{0pt}{1em}\textsuperscript{2}ESE: Empirical Standard Error}\\
\multicolumn{7}{l}{\rule{0pt}{1em}\textsuperscript{3}RMSE: Root Mean Square Error}\\
\end{tabular}
\end{table}


\begin{table}[!ht]\centering
\caption{\label{orsini:tab6}Performance of the frequentist and Bayesian methods on the estimation of the difference of restricted mean survival time (dRMST) between two treatment groups, adjusted for prognostic covariates $Z_j$ and non-prognostic covariates $X_j$, with non-proportional hazards treatment effect (scenario 4: early effect true dRMST, $\delta = 0.9532$), with $30\%$ of censoring and different sample sizes. The gray shaded lines represent the correctly specified model.}
\medskip
    \scalebox{0.62}{
\begin{tabular}{lllrrrrr}
\toprule
n  & Methods & Adjustment variable & Bias & ASE$^1$ & ESE$^2$ & RMSE$^3$ & Coverage\\
\midrule
\addlinespace[0.3em]
50&\multicolumn{5}{l}{\textbf{Frequentist}}\\
&\hspace{1em}KM estimator & - & -0.0383 & 0.496 & 0.523 & 0.524 & 93.4\\
&\hspace{1em}Zucker (1998) & - & -0.0364 & 0.514 & 0.542 & 0.543 & 93.4\\
\rowcolor{gray!25}&\hspace{1em}Zucker (1998) & $\rm{\rm{Z_1}}$ & -0.0339 & 0.481 & 0.513 & 0.514 & 92.4\\
&\hspace{1em}Zucker (1998) & $\rm{\rm{Z_1}}$, $\rm{X_1}$ & -0.0461 & 0.483 & 0.515 & 0.517 & 92.7\\
&\hspace{1em}Zucker (1998) & $\rm{\rm{Z_1}}$, $\rm{X_1}$, $\rm{X_2}$, $\rm{X_3}$ & -0.0424 & 0.488 & 0.530 & 0.532 & 92.7\\
&\hspace{1em}Andersen et al. (2004) & - & -0.0379 & 0.501 & 0.525 & 0.526 & 93.4\\
\rowcolor{gray!25}&\hspace{1em}Andersen et al. (2004) & $\rm{\rm{Z_1}}$ & -0.0328 & 0.479 & 0.499 & 0.500 & 92.4\\
&\hspace{1em}Andersen et al. (2004) & $\rm{\rm{Z_1}}$, $\rm{X_1}$ & -0.0347 & 0.481 & 0.504 & 0.505 & 92.4\\
&\hspace{1em}Andersen et al. (2004) & $\rm{\rm{Z_1}}$, $\rm{X_1}$, $\rm{X_2}$, $\rm{X_3}$ & -0.0291 & 0.485 & 0.520 & 0.520 & 92.5\\
\addlinespace[0.3em]
&\multicolumn{5}{l}{\textbf{Bayesian}}\\
&\hspace{1em}Zhang and Yin (2023) & - & -0.0138 & 0.491 & 0.563 & 0.564 & 90.8\\
&\hspace{1em}GMM & - & -0.0340 & 0.517 & 0.503 & 0.504 & 94.5\\
\rowcolor{gray!25}&\hspace{1em}GMM & $\rm{\rm{Z_1}}$ & -0.0086 & 0.532 & 0.486 & 0.486 & 95.5\\
&\hspace{1em}GMM & $\rm{\rm{Z_1}}$, $\rm{X_1}$ & -0.0174 & 0.778 & 0.510 & 0.510 & 99.4\\
&\hspace{1em}GMM  & $\rm{\rm{Z_1}}$, $\rm{X_1}$, $\rm{X_2}$, $\rm{X_3}$ & -0.1776 & 1.666 & 0.612 & 0.638 & 100.0\\
\midrule
\addlinespace[0.3em]
100&\multicolumn{5}{l}{\textbf{Frequentist}}\\
&\hspace{1em}KM estimator & - & 0.0072 & 0.361 & 0.357 & 0.357 & 94.3\\
&\hspace{1em}Zucker (1998) & - & -0.0001 & 0.364 & 0.356 & 0.356 & 94.9\\
\rowcolor{gray!25}&\hspace{1em}Zucker (1998) & $\rm{\rm{Z_1}}$ & -0.0111 & 0.339 & 0.329 & 0.329 & 93.7\\
&\hspace{1em}Zucker (1998) & $\rm{\rm{Z_1}}$, $\rm{X_1}$ & -0.0113 & 0.339 & 0.333 & 0.333 & 93.6\\
&\hspace{1em}Zucker (1998) & $\rm{\rm{Z_1}}$, $\rm{X_1}$, $\rm{X_2}$, $\rm{X_3}$ & -0.0100 & 0.339 & 0.340 & 0.340 & 92.9\\
&\hspace{1em}Andersen et al. (2004) & - & 0.0071 & 0.363 & 0.358 & 0.358 & 94.8\\
\rowcolor{gray!25}&\hspace{1em}Andersen et al. (2004) & $\rm{\rm{Z_1}}$ & 0.0003 & 0.346 & 0.339 & 0.339 & 94.5\\
&\hspace{1em}Andersen et al. (2004) & $\rm{\rm{Z_1}}$, $\rm{X_1}$ & 0.0013 & 0.346 & 0.342 & 0.342 & 94.5\\
&\hspace{1em}Andersen et al. (2004) & $\rm{\rm{Z_1}}$, $\rm{X_1}$, $\rm{X_2}$, $\rm{X_3}$ & 0.0018 & 0.346 & 0.345 & 0.345 & 93.9\\

\addlinespace[0.3em]
&\multicolumn{5}{l}{\textbf{Bayesian}}\\
&\hspace{1em}Zhang and Yin (2023) & - & 0.0103 & 0.357 & 0.361 & 0.361 & 93.8\\
&\hspace{1em}GMM & - & 0.0063 & 0.367 & 0.351 & 0.351 & 95.3\\
\rowcolor{gray!25}&\hspace{1em}GMM & $\rm{\rm{Z_1}}$ & 0.0122 & 0.358 & 0.334 & 0.335 & 95.3\\
&\hspace{1em}GMM & $\rm{\rm{Z_1}}$, $\rm{X_1}$ & 0.0151 & 0.372 & 0.337 & 0.337 & 95.4\\
&\hspace{1em}GMM & $\rm{\rm{Z_1}}$, $\rm{X_1}$, $\rm{X_2}$, $\rm{X_3}$ & 0.0174 & 0.413 & 0.345 & 0.345 & 96.5\\
\midrule
\addlinespace[0.3em]
200&\multicolumn{7}{l}{\textbf{Frequentist}}\\
&\hspace{1em}KM estimator & - & -0.0056 & 0.257 & 0.266 & 0.266 & 93.8\\
&\hspace{1em}Zucker (1998) & - & -0.0104 & 0.258 & 0.264 & 0.264 & 93.8\\
\rowcolor{gray!25}&\hspace{1em}Zucker (1998) & $\rm{\rm{Z_1}}$ & -0.0133 & 0.239 & 0.243 & 0.243 & 93.9\\
&\hspace{1em}Zucker (1998) & $\rm{Z_1}$, $\rm{X_1}$ & -0.0128 & 0.239 & 0.245 & 0.245 & 93.9\\
&\hspace{1em}Zucker (1998) & $\rm{Z_1}$, $\rm{X_1}$, $\rm{X_2}$, $\rm{X_3}$ & -0.0117 & 0.239 & 0.248 & 0.248 & 93.8\\
&\hspace{1em}Andersen et al. (2004) & - & -0.0056 & 0.258 & 0.266 & 0.266 & 93.8\\
\rowcolor{gray!25}&\hspace{1em}Andersen et al. (2004) & $\rm{Z_1}$  & -0.0088 & 0.246 & 0.251 & 0.251 & 93.9\\
&\hspace{1em}Andersen et al. (2004) & $\rm{Z_1}$, $\rm{X_1}$ & -0.0074 & 0.246 & 0.251 & 0.251 & 93.9\\
&\hspace{1em}Andersen et al. (2004) & $\rm{Z_1}$, $\rm{X_1}$, $\rm{X_2}$, $\rm{X_3}$ & -0.0068 & 0.246 & 0.254 & 0.254 & 93.7\\
\addlinespace[0.3em]
&\multicolumn{7}{l}{\textbf{Bayesian}}\\
&\hspace{1em}Zhang and Yin (2023) & - & -0.0058 & 0.256 & 0.266 & 0.266 & 93.8\\
&\hspace{1em}GMM & - & -0.0070 & 0.259 & 0.263 & 0.263 & 94.5\\
\rowcolor{gray!25}&\hspace{1em}GMM & $\rm{Z_1}$ & -0.0033 & 0.250 & 0.249 & 0.249 & 94.6\\
&\hspace{1em}GMM & $\rm{Z_1}$, $\rm{X_1}$ & -0.0014 & 0.253 & 0.249 & 0.249 & 94.3\\
&\hspace{1em}GMM & $\rm{Z_1}$, $\rm{X_1}$, $\rm{X_2}$, $\rm{X_3}$ & -0.0008 & 0.261 & 0.252 & 0.252 & 95.3\\
\midrule
\addlinespace[0.3em]
500&\multicolumn{5}{l}{\textbf{Frequentist}}\\
&\hspace{1em}KM estimator & - & 0.0055 & 0.163 & 0.167 & 0.167 & 93.9\\
&\hspace{1em}Zucker (1998) & - & 0.0036 & 0.163 & 0.167 & 0.167 & 94.1\\
\rowcolor{gray!25}&\hspace{1em}Zucker (1998) & $\rm{Z_1}$ & 0.0011 & 0.151 & 0.154 & 0.154 & 94.6\\
&\hspace{1em}Zucker (1998) & $\rm{Z_1}$, $\rm{X_1}$ & 0.0010 & 0.151 & 0.154 & 0.154 & 94.4\\
&\hspace{1em}Zucker (1998) & $\rm{Z_1}$, $\rm{X_1}$, $\rm{X_2}$, $\rm{X_3}$ & 0.0012 & 0.151 & 0.155 & 0.155 & 94.4\\
&\hspace{1em}Andersen et al. (2004) & - & 0.0055 & 0.163 & 0.167 & 0.167 & 93.9\\
\rowcolor{gray!25}&\hspace{1em}Andersen et al. (2004) & $\rm{Z_1}$ & 0.0037 & 0.156 & 0.159 & 0.159 & 94.9\\
&\hspace{1em}Andersen et al. (2004) & $\rm{Z_1}$, $\rm{X_1}$ & 0.0035 & 0.156 & 0.159 & 0.159 & 94.7\\
&\hspace{1em}Andersen et al. (2004) & $\rm{Z_1}$, $\rm{X_1}$, $\rm{X_2}$, $\rm{X_3}$ & 0.0040 & 0.155 & 0.159 & 0.159 & 94.5\\
\addlinespace[0.3em]
&\multicolumn{5}{l}{\textbf{Bayesian}}\\
&\hspace{1em}Zhang and Yin (2023) & - & 0.0056 & 0.162 & 0.167 & 0.167 & 93.9\\
&\hspace{1em}GMM & - & 0.0044 & 0.163 & 0.166 & 0.166 & 94.2\\
\rowcolor{gray!25}&\hspace{1em}GMM & $\rm{Z_1}$ & 0.0061 & 0.156 & 0.158 & 0.158 & 94.9\\
&\hspace{1em}GMM & $\rm{Z_1}$, $\rm{X_1}$ & 0.0062 & 0.157 & 0.159 & 0.159 & 94.4\\
&\hspace{1em}GMM & $\rm{Z_1}$, $\rm{X_1}$, $\rm{X_2}$, $\rm{X_3}$ & 0.0067 & 0.159 & 0.159 & 0.159 & 95.3\\
\bottomrule
\multicolumn{8}{l}{\rule{0pt}{1em}\textsuperscript{1}ASE: Average Standard Error,\textsuperscript{2}ESE: Empirical Standard Error,\textsuperscript{3}RMSE: Root Mean Square Error}\\
\multicolumn{8}{l}{\rule{0pt}{1em} Prognostic variable $\rm{Z_1} \sim U(0,2)$, other variables $\rm{X_1} \sim N(0,1)$, $\rm{X_2} \sim Bin(0.5)$, $\rm{X_3} \sim U(0,2)$}\\
\end{tabular}
}
\end{table}


\begin{table}[!ht]\centering
\caption{\label{orsini:tab7}Performance of the methods on the estimation of the dRMST between two treatment groups, adjusted for prognostic covariates $Z_j$ and non-prognostic covariates $X_j$, with non-proportional hazards treatment effect (scenario 5: late effect true dRMST, $\delta = 0.4911$), with $30\%$ of censoring and different sample sizes. The gray shaded lines represent the correctly specified model.}
\medskip
 \scalebox{0.62}{
\begin{tabular}{lllrrrrr}
\toprule
n & Methods & Adjustment variable & Bias & ASE$^1$ & ESE$^2$ & RMSE$^3$ & Coverage\\
\midrule
\addlinespace[0.3em]
50&\multicolumn{5}{l}{\textbf{Frequentist}}\\
&\hspace{1em}KM estimator & - & -0.0245 & 0.496 & 0.500 & 0.501 & 94.7\\
&\hspace{1em}Zucker (1998) & - & -0.0228 & 0.499 & 0.499 & 0.500 & 94.8\\
&\hspace{1em}Zucker (1998) & $\rm{Z_1}$ & -0.0074 & 0.433 & 0.433 & 0.433 & 94.2\\
\rowcolor{gray!25}&\hspace{1em}Zucker (1998) & $\rm{Z_1}$, $\rm{Z_2}$ & -0.0085 & 0.426 & 0.433 & 0.434 & 94.0\\
&\hspace{1em}Zucker (1998) & $\rm{Z_1}$, $\rm{Z_2}$, $\rm{X_1}$, $\rm{X_2}$ & -0.0030 & 0.427 & 0.451 & 0.451 & 92.2\\
&\hspace{1em}Andersen et al. (2004) & - & -0.0247 & 0.499 & 0.501 & 0.501 & 94.8\\
&\hspace{1em}Andersen et al. (2004) & $\rm{Z_1}$ & -0.0063 & 0.452 & 0.457 & 0.457 & 94.1\\
\rowcolor{gray!25}&\hspace{1em}Andersen et al. (2004) & $\rm{Z_1}$, $\rm{Z_2}$ & -0.0088 & 0.448 & 0.466 & 0.466 & 93.1\\
&\hspace{1em}Andersen et al. (2004) & $\rm{Z_1}$, $\rm{Z_2}$, $\rm{X_1}$, $\rm{X_2}$& -0.0050 & 0.446 & 0.476 & 0.476 & 91.9\\
\addlinespace[0.3em]
&\multicolumn{6}{l}{\textbf{Bayesian}}\\
&\hspace{1em}Zhang and Yin (2023) & - & -0.0064 & 0.486 & 0.515 & 0.515 & 94.2\\
&\hspace{1em}GMM & - & -0.0049 & 0.517 & 0.484 & 0.484 & 96.2\\
&\hspace{1em}GMM & $\rm{Z_1}$ & 0.0307 & 0.568 & 0.454 & 0.455 & 97.5\\
\rowcolor{gray!25}&\hspace{1em}GMM & $\rm{Z_1}$, $\rm{Z_2}$ & 0.0547 & 0.742 & 0.461 & 0.464 & 99.1\\
&\hspace{1em}GMM& $\rm{Z_1}$, $\rm{Z_2}$, $\rm{X_1}$, $\rm{X_2}$ & 0.1659 & 2.091 & 0.586 & 0.609 & 100.0\\
\midrule
\addlinespace[0.3em]
100&\multicolumn{5}{l}{\textbf{Frequentist}}\\
&\hspace{1em}KM estimator & - & -0.0042 & 0.353 & 0.338 & 0.338 & 95.3\\
&\hspace{1em}Zucker (1998) & - & -0.0107 & 0.353 & 0.334 & 0.334 & 95.6\\
&\hspace{1em}Zucker (1998) & $\rm{Z_1}$ & -0.0151 & 0.304 & 0.301 & 0.301 & 94.6\\
\rowcolor{gray!25}&\hspace{1em}Zucker (1998) & $\rm{Z_1}$, $\rm{Z_2}$ & -0.0160 & 0.300 & 0.299 & 0.299 & 93.7\\
&\hspace{1em}Zucker (1998) & $\rm{Z_1}$, $\rm{Z_2}$, $\rm{X_1}$, $\rm{X_2}$ & -0.0127 & 0.299 & 0.301 & 0.301 & 93.6\\
&\hspace{1em}Andersen et al. (2004) & - & -0.0041 & 0.355 & 0.338 & 0.338 & 95.8\\
&\hspace{1em}Andersen et al. (2004) & $\rm{Z_1}$ & -0.0070 & 0.321 & 0.310 & 0.310 & 95.4\\
\rowcolor{gray!25}&\hspace{1em}Andersen et al. (2004) & $\rm{Z_1}$, $\rm{Z_2}$ & -0.0063 & 0.318 & 0.309 & 0.309 & 94.7\\
&\hspace{1em}Andersen et al. (2004) & $\rm{Z_1}$, $\rm{Z_2}$, $\rm{X_1}$, $\rm{X_2}$ & -0.0061 & 0.318 & 0.307 & 0.307 & 94.5\\
\addlinespace[0.3em]
&\multicolumn{6}{l}{\textbf{Bayesian}}\\
&\hspace{1em}Zhang and Yin (2023) & - & -0.0035 & 0.349 & 0.338 & 0.338 & 95.5\\
&\hspace{1em}GMM & - & 0.0032 & 0.359 & 0.333 & 0.333 & 96.4\\
&\hspace{1em}GMM & $\rm{Z_1}$ & 0.0070 & 0.337 & 0.307 & 0.307 & 96.5\\
\rowcolor{gray!25}&\hspace{1em}GMM & $\rm{Z_1}$, $\rm{Z_2}$ & 0.0079 & 0.341 & 0.305 & 0.306 & 96.6\\
&\hspace{1em}GMM & $\rm{Z_1}$, $\rm{Z_2}$, $\rm{X_1}$, $\rm{X_2}$& 0.0105 & 0.415 & 0.305 & 0.305 & 98.2\\
\midrule
\addlinespace[0.3em]
200&\multicolumn{5}{l}{\textbf{Frequentist}}\\
&\hspace{1em}KM estimator & - & -0.0062 & 0.251 & 0.262 & 0.263 & 93.5\\
&\hspace{1em}Zucker (1998) & - & -0.0098 & 0.251 & 0.261 & 0.261 & 93.5\\
&\hspace{1em}Zucker (1998) & $\rm{Z_1}$ & -0.0024 & 0.215 & 0.223 & 0.223 & 94.6\\
\rowcolor{gray!25}&\hspace{1em}Zucker (1998) & $\rm{Z_1}$, $\rm{Z_2}$ & -0.0031 & 0.211 & 0.219 & 0.219 & 94.7\\
&\hspace{1em}Zucker (1998) & $\rm{Z_1}$, $\rm{Z_2}$, $\rm{X_1}$, $\rm{X_2}$ & -0.0031 & 0.211 & 0.220 & 0.220 & 94.0\\
&\hspace{1em}Andersen et al. (2004) & - & -0.0061 & 0.252 & 0.263 & 0.263 & 93.5\\
&\hspace{1em}Andersen et al. (2004) & $\rm{Z_1}$ & -0.0025 & 0.227 & 0.235 & 0.235 & 93.8\\
\rowcolor{gray!25}&\hspace{1em}Andersen et al. (2004) & $\rm{Z_1}$, $\rm{Z_2}$ & -0.0045 & 0.225 & 0.233 & 0.233 & 93.9\\
&\hspace{1em}Andersen et al. (2004) & $\rm{Z_1}$, $\rm{Z_2}$, $\rm{X_1}$, $\rm{X_2}$ & -0.0039 & 0.225 & 0.234 & 0.234 & 93.8\\
\addlinespace[0.3em]
&\multicolumn{6}{l}{\textbf{Bayesian}}\\
&\hspace{1em}Zhang and Yin (2023) & - & -0.0061 & 0.250 & 0.262 & 0.263 & 93.6\\
&\hspace{1em}GMM & - & -0.0029 & 0.253 & 0.261 & 0.261 & 94.2\\
&\hspace{1em}GMM & $\rm{Z_1}$ & 0.0046 & 0.233 & 0.234 & 0.234 & 94.8\\
\rowcolor{gray!25}&\hspace{1em}GMM & $\rm{Z_1}$, $\rm{Z_2}$ & 0.0021 & 0.232 & 0.232 & 0.232 & 94.7\\
&\hspace{1em}GMM & $\rm{Z_1}$, $\rm{Z_2}$, $\rm{X_1}$, $\rm{X_2}$ & 0.0032 & 0.239 & 0.234 & 0.234 & 95.2\\
\midrule
\addlinespace[0.3em]
500&\multicolumn{5}{l}{\textbf{Frequentist}}\\
&\hspace{1em}KM estimator & - & 0.0019 & 0.159 & 0.161 & 0.161 & 94.8\\
&\hspace{1em}Zucker (1998) & - & 0.0004 & 0.159 & 0.161 & 0.161 & 94.9\\
&\hspace{1em}Zucker (1998) & $\rm{Z_1}$ & 0.0028 & 0.136 & 0.139 & 0.139 & 94.6\\
\rowcolor{gray!25}&\hspace{1em}Zucker (1998) & $\rm{Z_1}$, $\rm{Z_2}$ & 0.0031 & 0.134 & 0.135 & 0.136 & 94.9\\
&\hspace{1em}Zucker (1998) & $\rm{Z_1}$, $\rm{Z_2}$, $\rm{X_1}$, $\rm{X_2}$ & 0.0034 & 0.134 & 0.135 & 0.136 & 95.1\\
&\hspace{1em}Andersen et al. (2004) & - & 0.0018 & 0.160 & 0.161 & 0.161 & 94.8\\
&\hspace{1em}Andersen et al. (2004) & $\rm{Z_1}$ & 0.0045 & 0.144 & 0.147 & 0.147 & 95.3\\
\rowcolor{gray!25}&\hspace{1em}Andersen et al. (2004) & $\rm{Z_1}$, $\rm{Z_2}$ & 0.0041 & 0.143 & 0.145 & 0.145 & 95.2\\
&\hspace{1em}Andersen et al. (2004) & $\rm{Z_1}$, $\rm{Z_2}$, $\rm{X_1}$, $\rm{X_2}$& 0.0043 & 0.143 & 0.145 & 0.145 & 95.4\\
\addlinespace[0.3em]
&\multicolumn{6}{l}{\textbf{Bayesian}}\\
&\hspace{1em}Zhang and Yin (2023) & - & 0.0018 & 0.159 & 0.161 & 0.161 & 94.6\\
&\hspace{1em}GMM & - & 0.0028 & 0.160 & 0.161 & 0.161 & 94.8\\
&\hspace{1em}GMM & $\rm{Z_1}$ & 0.0078 & 0.146 & 0.146 & 0.147 & 95.5\\
\rowcolor{gray!25}&\hspace{1em}GMM & $\rm{Z_1}$, $\rm{Z_2}$ & 0.0068 & 0.144 & 0.145 & 0.145 & 95.5\\
&\hspace{1em}GMM & $\rm{Z_1}$, $\rm{Z_2}$, $\rm{X_1}$, $\rm{X_2}$& 0.0071 & 0.146 & 0.145 & 0.145 & 96.3\\
\bottomrule
\multicolumn{8}{l}{\rule{0pt}{1em}\textsuperscript{1}ASE: Average Standard Error,\textsuperscript{2}ESE: Empirical Standard Error,\textsuperscript{3}RMSE: Root Mean Square Error}\\
\multicolumn{8}{l}{\rule{0pt}{1em}Prognostic variables $\rm{Z_1} \sim N(0,1)$ and $\rm{Z_2} \sim Bin(0.5)$, other variables $\rm{X_1} \sim N(0,1)$, $\rm{X_2} \sim Bin(0.5)$, and $\rm{X_3} \sim U(0,2)$}\\
\end{tabular}
}
\end{table}

\begin{table}[!ht]\centering

\caption{\label{orsini:tab8}Performance of the frequentist and Bayesian methods on the univariate estimation of the difference of restricted mean survival time (dRMST) between two treatment groups with non-proportional hazards treatment effect (scenario 6: crossing hazards true dRMST, $\delta = -0.1258$), with $30\%$ of censoring and different sample sizes.}
\medskip
\begin{tabular}{llrrrrr}
\toprule
n & Methods & Bias & ASE$^1$& ESE$^2$  & RMSE$^3$ & Coverage\\
\midrule
\addlinespace[0.3em]
50&\multicolumn{5}{l}{\textbf{Frequentist}}\\
&\hspace{1em}KM estimator & 0.0225 & 0.505 & 0.544 & 0.544 & 92.1\\
&\hspace{1em}Zucker (1998) & 0.0249 & 0.500 & 0.532 & 0.533 & 92.7\\
&\hspace{1em}Andersen et al. (2004) & 0.0224 & 0.505 & 0.543 & 0.544 & 92.4\\
\addlinespace[0.3em]
&\multicolumn{6}{l}{\textbf{Bayesian}}\\
&\hspace{1em}Zhang and Yin (2023) & 0.0227 & 0.494 & 0.544 & 0.545 & 92.0\\
&\hspace{1em}GMM & 0.0636 & 0.523 & 0.528 & 0.532 & 94.1\\
\midrule
\addlinespace[0.3em]
100&\multicolumn{5}{l}{\textbf{Frequentist}}\\
&\hspace{1em}KM estimator & -0.0142 & 0.360 & 0.359 & 0.360 & 94.5\\
&\hspace{1em}Zucker (1998) & -0.0128 & 0.358 & 0.356 & 0.356 & 94.6\\
&\hspace{1em}Andersen et al. (2004) & -0.0142 & 0.360 & 0.359 & 0.359 & 94.5\\
\addlinespace[0.3em]
&\multicolumn{6}{l}{\textbf{Bayesian}}\\
&\hspace{1em}Zhang and Yin (2023) & -0.0140 & 0.356 & 0.360 & 0.360 & 94.8\\
&\hspace{1em}GMM & 0.0051 & 0.365 & 0.354 & 0.354 & 95.3\\
\midrule
\addlinespace[0.3em]
200&\multicolumn{5}{l}{\textbf{Frequentist}}\\
&\hspace{1em}KM estimator & 0.0080 & 0.255 & 0.257 & 0.257 & 94.7\\
&\hspace{1em}Zucker (1998) & 0.0085 & 0.254 & 0.255 & 0.255 & 94.9\\
&\hspace{1em}Andersen et al. (2004) & 0.0080 & 0.255 & 0.257 & 0.257 & 94.7\\
\addlinespace[0.3em]
&\multicolumn{6}{l}{\textbf{Bayesian}}\\
&\hspace{1em}Zhang and Yin (2023) & 0.0080 & 0.254 & 0.257 & 0.257 & 94.9\\
&\hspace{1em}GMM & 0.0168 & 0.256 & 0.255 & 0.256 & 95.0\\
\midrule
\addlinespace[0.3em]
500&\multicolumn{5}{l}{\textbf{Frequentist}}\\
&\hspace{1em}KM estimator & -0.0034 & 0.162 & 0.166 & 0.166 & 93.8\\
&\hspace{1em}Zucker (1998) & -0.0032 & 0.161 & 0.165 & 0.165 & 93.8\\
&\hspace{1em}Andersen et al. (2004) & -0.0034 & 0.162 & 0.166 & 0.166 & 93.7\\
\addlinespace[0.3em]
&\multicolumn{6}{l}{\textbf{Bayesian}}\\
&\hspace{1em}Zhang and Yin (2023) & -0.0033 & 0.161 & 0.166 & 0.166 & 93.9\\
&\hspace{1em}GMM & -0.0003 & 0.162 & 0.165 & 0.165 & 94.1\\
\bottomrule
\multicolumn{6}{l}{\rule{0pt}{1em}\textsuperscript{1}ASE: Average Standard Error}\\
\multicolumn{7}{l}{\rule{0pt}{1em}\textsuperscript{2}ESE: Empirical Standard Error}\\
\multicolumn{6}{l}{\rule{0pt}{1em}\textsuperscript{3}RMSE: Root Mean Square Error}\\
\end{tabular}
\end{table}

\begin{table*}[!ht]\centering
\caption{\label{orsini:tab9}Performance of the frequentist and Bayesian methods, based on pseudo-observations, on the estimation of the difference of restricted mean survival time (dRMST) between two treatment groups, adjusted for a predictive covariate $\rm E$ and an interaction term $\rm A \times E$, with non-proportional hazards treatment effect (scenario 6: crossing hazards), with $30\%$ of censoring and different sample sizes. The parameter $\delta^- = - 0.9025$ corresponds to the true dRMST between the two treatment arms for patients with $\rm E = E^-$, $\delta^+ = 0.6492$ corresponds to the true dRMST between the two treatment arms for patients with $\rm E = E^+$, and $\beta_1 = 1.0644$ corresponds to the true dRMST between $\rm E^+$ and $\rm E^-$ in control patients.}
\medskip
 \scalebox{0.95}{
\begin{tabular}{cllrrrrr}
\toprule
n & Methods & Estimated coefficient & Bias & ASE$^1$ & ESE$^2$ & RMSE$^3$ & Coverage\\
\midrule
\addlinespace[0.3em]
50&\multicolumn{5}{l}{\textbf{Frequentist}}\\
&\hspace{1em}Andersen et al. (2004) & $\beta_1$  & -0.0172 & 0.650 & 0.671 & 0.671 & 92.4\\
&\hspace{1em}Andersen et al. (2004) & $\delta^{-}$ & 0.0108 & 0.604 & 0.635 & 0.635 & 92.9\\
&\hspace{1em}Andersen et al. (2004) & $\delta^{+}$ & 0.0206 & 0.574 & 0.615 & 0.615 & 92.6\\
\addlinespace[0.3em]
&\multicolumn{7}{l}{\textbf{Bayesian}}\\
&\hspace{1em}GMM & $\beta_1$ & 0.0534 & 0.784 & 0.598 & 0.600 & 97.0\\
&\hspace{1em}GMM & $\delta^{-}$ & 0.1255 & 0.808 & 0.570 & 0.583 & 98.4\\
&\hspace{1em}GMM & $\delta^{+}$ & -0.0014 & 0.813 & 0.555 & 0.555 & 98.0\\
\midrule
\addlinespace[0.3em]
100&\multicolumn{5}{l}{\textbf{Frequentist}}\\
&\hspace{1em}Andersen et al. (2004) & $\beta_1$  & 0.0171 & 0.463 & 0.472 & 0.473 & 94.1\\
&\hspace{1em}Andersen et al. (2004) & $\delta^{-}$ & -0.0080 & 0.431 & 0.434 & 0.434 & 94.3\\
&\hspace{1em}Andersen et al. (2004) & $\delta^{+}$ & -0.0115 & 0.412 & 0.417 & 0.418 & 94.2\\
\addlinespace[0.3em]
&\multicolumn{7}{l}{\textbf{Bayesian}}\\
&\hspace{1em}GMM & $\beta_1$ & 0.0455 & 0.492 & 0.453 & 0.455 & 96.1\\
&\hspace{1em}GMM & $\delta^{-}$ & 0.0329 & 0.462 & 0.420 & 0.421 & 96.4\\
&\hspace{1em}GMM & $\delta^{+}$ & -0.0106 & 0.443 & 0.407 & 0.408 & 96.4\\
\midrule
\addlinespace[0.3em]
200&\multicolumn{5}{l}{\textbf{Frequentist}}\\
&\hspace{1em}Andersen et al. (2004) & $\beta_1$ & 0.0074 & 0.330 & 0.340 & 0.340 & 93.9\\
&\hspace{1em}Andersen et al. (2004) & $\delta^{-}$ & 0.0060 & 0.307 & 0.310 & 0.310 & 94.0\\
&\hspace{1em}Andersen et al. (2004) & $\delta^{+}$ & 0.0021 & 0.292 & 0.309 & 0.309 & 92.4\\
\addlinespace[0.3em]
&\multicolumn{7}{l}{\textbf{Bayesian}}\\
&\hspace{1em}GMM & $\beta_1$ & 0.0212 & 0.339 & 0.333 & 0.334 & 95.2\\
&\hspace{1em}GMM & $\delta^{-}$ & 0.0258 & 0.315 & 0.305 & 0.306 & 94.6\\
&\hspace{1em}GMM & $\delta^{+}$ & 0.0022 & 0.301 & 0.306 & 0.306 & 93.8\\
\midrule
\addlinespace[0.3em]
500&\multicolumn{5}{l}{\textbf{Frequentist}}\\
&\hspace{1em}Andersen et al. (2004) & $\beta_1$  & 0.0047 & 0.209 & 0.216 & 0.216 & 94.4\\
&\hspace{1em}Andersen et al. (2004) & $\delta^{-}$ & -0.0008 & 0.194 & 0.205 & 0.205 & 93.9\\
&\hspace{1em}Andersen et al. (2004) & $\delta^{+}$ & -0.0029 & 0.186 & 0.195 & 0.195 & 93.7\\
\addlinespace[0.3em]
&\multicolumn{7}{l}{\textbf{Bayesian}}\\
&\hspace{1em}GMM & $\beta_1$ & 0.0104 & 0.212 & 0.214 & 0.215 & 94.7\\
&\hspace{1em}GMM & $\delta^{-}$ & 0.0074 & 0.196 & 0.203 & 0.203 & 93.9\\
&\hspace{1em}GMM & $\delta^{+}$ & -0.0029 & 0.189 & 0.195 & 0.195 & 94.3\\
\bottomrule
\multicolumn{8}{l}{\rule{0pt}{1em}\textsuperscript{1}ASE: Average Standard Error,\textsuperscript{2}ESE: Empirical Standard Error,\textsuperscript{3}RMSE: Root Mean Square Error}\\
\end{tabular}
}
\vspace{1em}
\end{table*}


  \begin{table}[!ht]\centering
          \caption {\label{orsini:tab10} Estimations of all parameters from the Bayesian generalized method of moments apply to the Getug-AFU 15 trial for the PSA progression-free survival analysis. The intercept coefficient represents the RMST for the control group, with all binary covariates set to their reference levels. The remaining coefficients represent the dRMST between the two groups defined by each corresponding binary explanatory variable in the model.}
\begin{tabular}{lrrrrrrr}
\toprule
  & & &  \multicolumn{5}{c}{Quantiles}\\
  \cmidrule(lr){2-3}  \cmidrule(lr){4-8}
  Covariates  & $\widehat\beta^1$ & SE$^2$ &  2.5\% & 25\% & 50\% & 75\% & 97.5\%\\
\midrule
Intercept ($\beta_0$) & 5.60 &0.54&4.52&  5.22&  5.60 & 5.96&  6.67\\
Gleason score ($\beta_1$) &-0.19 & 0.17 &-0.53 & -0.31 &-0.19& -0.08 & 0.15\\
ECOG performance status ($\beta_2$) & -0.55&0.18 &-0.92 &-0.68& -0.56& -0.43 &-0.19\\
Alkaline phosphatase concentration ($\beta_3$) &-1.25 & 0.19  & -1.63& -1.38& -1.25& -1.13 &-0.87\\
Presence of bone metastases ($\beta_4$) & -0.46&0.27&-1.00 &-0.64 &-0.46& -0.28& 0.05\\
Treatment ($\delta$) & 0.57&0.17&0.24&  0.45 & 0.57 & 0.68 & 0.91\\
\bottomrule
\multicolumn{8}{l}{\rule{0pt}{1em} \textsuperscript{1}$\widehat\beta$: estimated regression coefficient \textsuperscript{2}SE: Standard Error}
\end{tabular}
    \end{table}


\begin{sidewaystable}
\caption{\label{orsini:tab11} Unadjusted estimation of the difference of $\tau$-RMST between the two treatment arms and $95\%$ confidence or credibility intervals (CI) from the Getug-AFU 15 trial for different values of the restricted time $\tau$.} 
\begin{tabular}{lrlrrr}
\toprule
Choice of $\tau$ & $\tau$ & Methods & $\widehat{\delta}$ & CI (low) & CI (up) \\
\midrule
\color{black}{(a) 90th percentile of the observed times} & 4.599 & KM estimator & 0.555 & 0.246 & 0.924 \\
& & Zucker (1998) & 0.553&  0.214 & 0.892\\
& & Andersen et al. (2004) & 0.555 & 0.216 & 0.894 \\
& & Zhang and Yin (2023) & 0.560 & 0.216 & 0.882\\
& & Bayesian GMM & 0.591 & 0.233 & 0.943 \\
\cmidrule(rl){3-6}
\color{black} (b1$^1$) SE(S(t)) $\leq$ 5\% & 4.608 & KM estimator & 0.556 & 0.246 & 0.925 \\
& & Zucker (1998) & 0.554& 0.214& 0.894\\
& & Andersen et al. (2004) & 0.556 & 0.216 & 0.895 \\
& & Zhang and Yin (2023) & 0.557 & 0.221 & 0.899\\
& & Bayesian GMM & 0.564 & 0.225 & 0.900 \\
\cmidrule(rl){3-6}
\color{black} (reference) 5 years & 5.000 & KM estimator & 0.585 & 0.216 & 0.955 \\
& & Zucker (1998) & 0.583 & 0.213 & 0.953\\
& & Andersen et al. (2004) & 0.585 & 0.215 & 0.955 \\
& & Zhang and Yin (2023) & 0.589 & 0.219 & 0.959\\
& & Bayesian GMM & 0.591 & 0.233 & 0.943 \\
\cmidrule(rl){3-6}
\color{black} (b2$^1$) SE(S(t))$\leq$7.5\% & 5.766 & KM estimator & 0.623 & 0.158 & 1.013 \\
& & Zucker (1998) & 0.621 & 0.192 &1.050\\
& & Andersen et al. (2004) & 0.621 & 0.193 & 1.050 \\
& & Zhang and Yin (2023) & 0.624 & 0.173 & 1.053\\
& & Bayesian GMM & 0.619 & 0.179 & 1.075 \\
\cmidrule(rl){3-6}
(c) Min of the max observed time in each group & 6.658 & KM estimator & 0.603 & 0.096 & 1.075 \\
& & Zucker (1998) &0.608& 0.115& 1.099\\
& & Andersen et al. (2004) & 0.608 & 0.115 & 1.100 \\
& & Zhang and Yin (2023) & 0.590 & 0.215 & 0.959\\
& & Bayesian GMM & 0.613 & 0.111 & 1.094 \\
\bottomrule
\multicolumn{6}{l}{$\rule{0pt}{1em}\textsuperscript{1}$(b) the largest time point $\rm t$ for which the standard error of the survival estimate is lower than a reasonable limit,}\\
\multicolumn{6}{l}{either $5\%$ (b1) or $7.5\%$ (b2)}
\end{tabular}
\end{sidewaystable}